\documentclass[paper,twocolumn,amsmath,amssymb,floatfix]{revtex4}
\usepackage[latin1]{inputenc}
\usepackage{graphicx}
\usepackage{pstricks}
\usepackage{bm}
\usepackage{amsmath,amssymb}
\usepackage{dcolumn}
\usepackage{bm}
\usepackage{epsfig}
\usepackage{epstopdf}
\usepackage{subfigure}
\usepackage{multirow}
\usepackage{feynmp}

\newcommand{\bfr}{ {\bf r}} 
\newcommand{\bfrp}{ {\bf r'}} 

\newcommand{\refeq}[1]{(\ref{#1})} 
\newcommand{\refcite}[1]{Ref.~\cite{#1}} 
\newcommand{\Fref}[1]{Fig.~\ref{#1}} 
\newcommand{\refsec}[1]{Sec.~\ref{#1}} 

\begin{document}

\title{Random-phase approximation and its applications in computational chemistry and materials science}
\author{Xinguo Ren,$^1$ Patrick Rinke,$^1$ Christian Joas,$^{1,2}$ and Matthias Scheffler$^1$}
\affiliation{$^1$Fritz-Haber-Institut der Max-Planck-Gesellschaft,
Faradayweg 4-6, 14195, Berlin, Germany \\
$^2$Max-Planck-Institut f{\"u}r Wissenschaftsgeschichte, Boltzmannstr. 22, 14195, Berlin, Germany}

\begin{abstract}
The random-phase approximation (RPA) as an approach for computing the electronic correlation
energy is reviewed. After a brief account of its basic concept and historical development, 
the paper is devoted to the theoretical formulations of RPA, and its applications to realistic systems. 
With several illustrating applications, we discuss the implications of RPA for computational 
chemistry and materials science. The computational cost of RPA is also addressed 
which is critical for its widespread use in future applications. In addition, current correction 
schemes going beyond RPA and directions of further development will be discussed.

\end{abstract}

\maketitle

\section{\label{sec:intro}Introduction}
	Computational materials science has developed into an indispensable discipline complementary to experimental 
materials science. The fundamental aim of computational materials science is to derive understanding entirely from the basic laws of physics, i.e., quantum mechanical first principles,  and increasingly also to make predictions of new properties or new materials for specific tasks. The rapid increase in available computer power together with new methodological developments are major factors in the growing impact of this field for practical applications to real materials. 

Density functional theory (DFT) \cite{Hohenberg/Kohn:1964} has shaped the realm of first principles materials science like no other method today. This success has been facilitated by the computational efficiency of the local-density \cite{Kohn/Sham:1965} or generalized gradient approximation \cite{Becke:1988a,Lee/Yang/Parr:1992,Perdew/Burke/Ernzerhof:1996} (LDA and GGA) of the exchange-correlation functional that make DFT applicable to polyatomic systems containing up to several thousand atoms. However, these approximations are subject to several well-known deficiencies. In the quest for finding an ``optimal" electronic structure method, that combines accuracy and tractability with transferability across different chemical environments and dimensionalities (e.g. molecules/clusters, wires/tubes, surfaces, solids) many new approaches, improvements and refinements have been proposed over the years. These have been classified by Perdew in his  ``Jacob's ladder" hierarchy \cite{Perdew/Schmidt:2001}. 

In this context, the treatment of exchange and correlation in terms of ``exact-exchange plus correlation in the random-phase approximation" \cite{Bohm/Pines:1953,Gell-Mann/Brueckner:1957} offers  a promising avenue. This is largely due to three attractive features. The exact-exchange energy cancels the spurious self-interaction error present in the Hartree energy exactly (although the RPA correlation itself does contain some ``self-correlation" and is non-zero for one-electron systems). The RPA correlation energy is fully non-local and includes long-range van der Waals (vdW) interactions automatically and seamlessly. Moreover, dynamic electronic screening is taken into account by summing up a sequence of ``ring" diagrams to infinite order, which makes RPA applicable to small-gap or metallic systems where finite-order many-body perturbation theories break down \cite{Gell-Mann/Brueckner:1957,Fetter/Walecka:1971,Grueneis/Marsman/Kresse:2010}. 

The random-phase approximation actually predates density-functional theory, but it took until the late 1970s to be formulated in the context of DFT \cite{Langreth/Perdew:1977} and until the early years of this millennium to be applied as a first principles electronic structure method \cite{Furche:2001,Fuchs/Gonze:2002}. We take the renewed and widespread 
interest of the RPA \cite{Furche:2001,Fuchs/Gonze:2002,Furche/Voorhis:2005,Scuseria/Henderson/Sorensen:2008,Janesko/Henderson/Scuseria:2009,Toulouse/etal:2009,Paier/etal:2010,Marini/Gonzalez/Rubio:2006,Jiang/Engel:2007,Harl/Kresse:2008,Harl/Kresse:2009,Lu/Li/Rocca/Galli:2009,Dobson/Wang:1999,Rohlfing/Bredow:2008,Ren/Rinke/Scheffler:2009,Schimka/etal:2010,Zhu/etal:2010,Lebegue/etal:2010,Eshuis/Yarkony/Furche:2010,Ismail-Beigi:2010,Goltl/Hafner:2011,Eshuis/Furche:2011,Ren/etal:2011,Hesselmann/Goerling:2011,Ruzsinszky/etal:2010,Ruzsinszky/etal:2011,Hesselmann:2011,Klopper/etal:2011,Eshuis/Bates/Furche:2011,Hellgren/etal:2012,Verma/Bartlett:2012} 
as motivation for this review article. To illustrate the unique development of this powerful physical concept, we will put the RPA into its historical context  before reviewing the basic theory. A summary of recent RPA results demonstrates the strength of this approach, but also its current limitations.  In addition we will discuss some of the most recent schemes going beyond RPA, including renormalized second order perturbation theory (r2PT), which is particularly promising in our opinion. We will also  address the issue of computational efficiency which, at present, impedes the widespread use of RPA, and indicate directions for further development.

\subsection{\label{sec:history}Early history}
	During the 1950s, quantum many-body theory underwent a major transformation, as concepts and techniques originating from quantum electrodynamics (QED)---in particular Feynman-Dyson diagrammatic perturbation theory---were extended to the study of solids and nuclei. A particularly important contribution at an early stage of this development was the RPA, a technique introduced by Bohm and Pines in a series of papers 
published in 1951--1953 \citep{Bohm:1951,Pines:1952,Bohm/Pines:1953,Pines:1953}. In recent years, the RPA has gained importance well beyond its initial realm of application, in computational condensed-matter physics, materials science, and quantum chemistry. 
While the RPA is commonly used within its diagrammatic formulation given by Gell-Mann and Brueckner \citep{Gell-Mann:1957}, it is nevertheless instructive to briefly discuss the history of its original formulation by Bohm and Pines. Some of the cited references are reprinted in \citep{Pines:1961}, which also recounts the history of the RPA until the early 1960s. Historical accounts of the work of Bohm and Pines can be found in Refs.~\citep{Hoddeson:1992_chapter8,Hughes:2006,Kojevnikov:2002} as well as in Refs.~\citep{Hesselmann/Goerling:2011,Eshuis/Bates/Furche:2011}.

In 1933--1934, Wigner and Seitz published two papers on the band structure of metallic sodium \citep{Wigner:1933,Wigner:1934a} in which they stressed the importance of the electronic correlation energy correction to band-theory calculations of the cohesive energy of metals. Wigner subsequently studied the interaction of electrons in a homogeneous electron gas (HEG) within a variational approach going beyond Hartree-Fock \citep{Wigner:1934b}. He initially provided estimates for the correlation energy only in the low- 
and high-density limits, but later interpolated to intermediate densities \citep{Wigner:1938}.
For almost two decades, Wigner's estimate remained the state of the art in the prototypical many-body problem of the HEG. According to a later statement by Herring,``the magnitude and role of correlation energy remained inadequately understood in a considerable part of the solid-state community for many years." \citep[pp.~71-72]{Herring:1980}

Due to the long-range nature of the Coulomb interaction and the resulting divergences, perturbative approaches, so successful in other areas, had to be complemented with approximations that accounted for the screening of the charge of an electron by the other electrons. Before the 1950s, these approximations commonly drew on work on classical electrolytes by Debye and H\"uckel \citep{Debye:1923a,Debye:1923b} and on work on heavy atoms by Thomas \citep{Thomas:1927} and Fermi \citep{Fermi:1927, Fermi:1928}, 
as well as later extensions \citep{Dirac:1930,Weizsacker:1935}.


As a reaction to work by Landsberg and Wohlfarth \citep{Landsberg:1949,Wohlfarth:1950}, Bohm and Pines in 1950 reported to have been led ``independently to the concept of an effective screened Coulomb force as a result of a systematical classical and quantum-mechanical investigation of the interaction of charges in an electron gas of high density'' \citep[p.~103]{Bohm:1950}. Their 1951--1953 series of papers \citep{Bohm:1951,Pines:1952,Bohm/Pines:1953,Pines:1953} presents this systematical investigation. The RPA was one of several physically-motivated approximations in the treatment of the HEG which allowed them to separate collective degrees of freedom (plasma oscillations) from single-particle degrees of freedom (which today would be called quasiparticles or charged excitations) via a suitable canonical transformation reminiscent of early work in QED \citep{Bloch:1937,Pauli:1938,Blum-Joas}. A similar theory was developed rather independently for nuclei by Bohr and Mottelson \citep{Bohr:1953}. 

In their first paper, illustrating the fundamental idea of separating single-particle and collective degrees of freedom, Bohm and Pines introduce RPA as one of four requirements \citep{Bohm:1951}:
\begin{quote}
``(3) We distinguish between two kinds of response of the electrons to a wave. One of these is in phase with the wave, so that the phase difference between the particle response and the wave producing it is independent of the position of the particle. This is the response which contributes to the organized behavior of the system. The other response has a phase difference with the wave producing it which depends on the position of the particle. Because of the general random location of the particles, this second response tends to average out to zero when we consider a large number of electrons, and we shall neglect the contributions arising from this. This procedure we call the {\it random phase approximation}.'' 
\end{quote}
In their second paper \citep{Pines:1952}, Bohm and Pines develop a detailed physical picture for the electronic behavior in a HEG due to the presence of Coulomb interactions. Only in their third paper, Bohm and Pines treat the (Coulomb-)interacting HEG quantum-mechanically \citep{Bohm/Pines:1953}. The RPA enables Bohm and Pines to absorb the long-range Coulomb interactions into the collective behavior of the system, leaving the single-particle degrees of freedom interacting only via a short-range screened force. 
The RPA amounts to neglecting the interaction between the collective and the single-particle degrees of freedom. Consequently, the momentum transfers of the Coulomb potential in Fourier space can be treated independently. The fourth paper \citep{Pines:1953} applies the new method to the electron gas in metals, discussing both validity and consequences of the RPA, such as the increase in electronic effective mass. 

%
Within condensed-matter theory, the significance of the Bohm-Pines approach quickly became apparent: Renormalizing the long-range Coulomb interaction into an effective screened interaction between new, effective single-particle degrees of freedom allowed both to overcome the divergences appearing in older theories of interacting many-body systems and to explain the hitherto puzzling success of the single-particle models of early  condensed-matter theory (see, e.g., \citep{Wigner:1938}). 
An early application of the RPA was Lindhard's calculation of 
the 
dielectric function of the electron gas \cite{Lindhard:1954}. 
Alternative approaches to and extensions of the Bohm-Pines approach were formulated by Tomonaga \citep{Tomonaga:1955a,Tomonaga:1955b}, and by Mott \citep{Mott:1954}, Fr\"ohlich and Pelzer \citep{Frohlich:1955}, and Hubbard \citep{Hubbard:1955a,Hubbard:1955b}.

In 1956, Landau's Fermi liquid theory \citep{Landau:1956} delivered the foundation for 
effective theories describing many-body systems in terms of quasiparticles. Brueckner \citep{Brueckner:1955} already in 1955 had introduced a ``linked-cluster expansion" for the treatment of nuclear matter (see also Ref.~\citep{Bethe:1956}). In 1957,  Goldstone \citep{Goldstone:1957}, using Feynman-like diagrams (based on Ref.~\citep{Gell-Mann:1951}), was able to show that Brueckner's theory is exact for the ground-state energy of an interacting many-fermion system.
This put the analogy between the QED vacuum and the ground state of a many-body system on firm ground. It had been introduced explicitly 
by Miyazawa for nuclei \citep{Miyazawa:1951} and by Salam for superconductors \citep{Salam:1953}, although the essence of the analogy dated back to the early days of quantum field theory in the 1930s. 

In late 1956, Gell-Mann and Brueckner employed a diagrammatic approach for treating the problem of the interacting electron gas. Their famous 1957 paper \citep{Gell-Mann:1957} eliminated the spurious divergences appearing in previous approaches. Expressing the perturbation series for the correlation energy of the HEG in terms of the Wigner-Seitz radius  $r_s$, they found that the divergences within earlier calculations  
\citep[e.g.,][]{Macke:1950} were mere artifacts: The logarithmic divergence appearing in the perturbative expansion of the correlation energy is canceled by similar divergences in higher-order terms. Summing the diagrams (which had a ring structure) to infinite order yielded a geometric series, and thus a convergent result. Gell-Mann and Brueckner derived an expression for the ground state energy of the interacting electron gas in the high-density limit. 
Their work, and Goldstone's paper \cite{Goldstone:1957}, are the earliest examples of the application of Feynman-type diagrammatic methods in condensed-matter theory.

Many applications of the new quantum-field theoretical methods followed: 
Gell-Mann calculated the specific heat of the high-density HEG 
\citep{Gell-Mann:1957b}; \citet{Hubbard:1957a,Hubbard:1957b} 
provided a description of the collective modes in terms of many-body perturbation theory (MBPT); Sawada {\it et al.} \citep{Sawada:1957,Sawada/etal:1957} demonstrated that the Gell-Mann-Brueckner approach indeed contained the plasma oscillations of Bohm and Pines \citep{Bohm/Pines:1953}, a point around which there had been quite some confusion initially \citep[see, e.g.,][]{Bohm:1957}. In addition, they demonstrated that the RPA is exact in the high-density limit. 
In 1958, Nozi\`eres and Pines formulated a many-body theory of the dielectric constant and showed the equivalence of Gell-Mann and Brueckner's diagrammatic approach and the RPA \citep{Nozieres:1958,Nozieres/Pines:1958}. 

\subsection{\label{sec:mod_history}RPA in modern times}
Today, the concept of RPA has gone far beyond
the domain of the HEG, and has gained considerable importance in computational physics 
and quantum chemistry. As a key example, RPA can be introduced within the framework
of DFT \cite{Hohenberg/Kohn:1964} via the so-called adiabatic-connection 
fluctuation-dissipation (ACFD) theorem \cite{Langreth/Perdew:1975,Langreth/Perdew:1977,Gunnarsson/Lundqvist:1976}.
Within this formulation, the unknown exact exchange-correlation (XC) energy functional in Kohn-Sham \cite{Kohn/Sham:1965}
DFT can be formally constructed by adiabatically switching on the Coulomb interaction between
electrons, while keeping the electron density fixed at its physical value.
This is formulated by a coupling-strength integration under which the integrand 
is related to the linear density-response function of fictitious systems with scaled Coulomb interaction.
Thus, an approximation to the response function directly translates into
an approximate DFT XC energy functional. RPA in this context is known
as an orbital-dependent energy functional \cite{Kuemmel/Kronik:2008} obtained by applying
the time-dependent Hartree approximation to the density response function.  

The versatility of RPA becomes apparent when considering alternative formulations. For instance, the RPA correlation energy may be understood as the shift of the zero-point plasmon excitation energies between the non-interacting and the fully interacting system, as shown by Sawada for the HEG \cite{Sawada/etal:1957}, and derived in detail by Furche \cite{Furche:2008} for general cases (see also Ref.~[\onlinecite{Ismail-Beigi:2010}]).   In quantum chemistry, RPA can also be interpreted as an approximation to coupled-cluster doubles (CCD) theory where only diagrams of ``ring" structure are kept \cite{Freeman:1977,Scuseria/Henderson/Sorensen:2008}. The equivalence of the ``plasmon" and ``ring-CCD formulation" of RPA has recently been established by Scuseria \textit{et al.} \cite{Scuseria/Henderson/Sorensen:2008}. These new perspectives not only offer more insight into the theory, but also help to devise more efficient algorithms to reduce the computational cost, e.g., by applying the 
Cholesky decomposition to the ``ring-CCD" equations \cite{Scuseria/Henderson/Sorensen:2008}. 

Following the early work on the HEG, other model electron systems were investigated, including the HEG surface  \cite{Pitarke/Eguiluz:1998,Kurth/Perdew:1999}, jellium slabs \cite{Dobson/Wang:1999}  and jellium spheres \cite{Beck:1984}.
The long-range behavior of RPA for spatially well-separated closed-shell subsystems was examined by Szabo and Ostlund
\cite{Szabo/Ostlund:1977}, as well as by Dobson \cite{Dobson:1994,Dobson/etal:2001,Dobson/Could:2012}. These authors showed that RPA 
yields the correct $1/R^6$ asymptotic behavior for the subsystem interaction. In addition, the long-range dispersion
interaction of RPA is fully consistent with the monomer polarizability computed at the same level of theory.
This is one of the main reasons for the revival of the RPA in recent years, because this long-range interaction is absent from LDA, GGA, and other popular density-functionals. Other reasons are the compatibility of the RPA correlation with exact exchange (which implies the exact cancellation of the self-interaction error present in the Hartree term) and the applicability to metallic systems.

For the HEG it has been demonstrated that RPA is not accurate for short-range correlation \cite{Hedin/Lundqvist:1969,Singwi/etal:1968}, and hence for a long time RPA was not considered to be valuable for realistic systems. Perdew and coworkers investigated this issue \cite{Kurth/Perdew:1999,Yan/Perdew/Kurth:2000}, and found that a local/semi-local correction to RPA has little effect on \textit{iso-electronic} energy differences, which suggests that  RPA might be accurate enough for many practical purposes. The application of RPA to realistic systems appeared slightly later, starting with  the pioneering work of Furche \cite{Furche:2001}, and Fuchs and Gonze \cite{Fuchs/Gonze:2002}  for small molecules. 
Accurate RPA total energies for closed-shell atoms were obtained by Jiang and Engel \cite{Jiang/Engel:2007}. Several groups investigated molecular properties, in particular in the weakly bound regime, with RPA and its variants \cite{Aryasetiawan/Miyake/Terakura:2002,Furche/Voorhis:2005,Toulouse/etal:2009,Zhu/etal:2010,Li/Lu/Nguyen/Galli:2010,Eshuis/Yarkony/Furche:2010,Ren/etal:2011,Eshuis/Furche:2011}, while others applied RPA to periodic systems \cite{Miyake/etal:2002,Marini/Gonzalez/Rubio:2006,Gonzalez/Fernandez/Rubio:2007,Harl/Kresse:2008,Lu/Li/Rocca/Galli:2009, Harl/Kresse:2009,Harl/Schimka/Kresse:2010}. Harl and Kresse in particular have performed extensive RPA benchmark studies for crystalline solids of all bonding types \cite{Harl/Kresse:2008,Harl/Kresse:2009,Harl/Schimka/Kresse:2010}.  At the same time, the application of RPA to surface adsorption problems has been reported \cite{Rohlfing/Bredow:2008,Ren/Rinke/Scheffler:2009,Schimka/etal:2010,Goltl/Hafner:2011,Ma/etal:2011,Kim/etal:2012}, with considerable success in resolving the ``CO adsorption puzzle".

Most practical RPA calculations in recent years have been performed non-self-consistently based on a preceding LDA or GGA reference calculation. In these calculations, the Coulomb integrals are usually not antisymmetrized in the evaluation of the RPA correlation energy, a practice sometimes called \emph{direct} RPA in the
quantum-chemical literature.  In this paper we will denote this common procedure ``standard RPA" to distinguish it from more sophisticated procedures. While a critical assessment of RPA is emerging and a wide variety of applications are pursued, certain shortcomings of  standard RPA have been noted. The most prominent is its systematic underestimation of binding energies \cite{Furche:2001,Ren/etal:2011,Harl/Kresse:2009}, and the failure to describe stretched radicals
\cite{Mori-Sanchez/etal:2012,Paier/etal:2010,Henderson/Scuseria:2010}. Over the years several attempts  have been made to improve upon RPA. The earliest is RPA+, where, as mentioned above, a local/semi-local correlation  correction based on LDA or GGA is added to the standard RPA correlation energy \cite{Kurth/Perdew:1999,Yan/Perdew/Kurth:2000}. Based on the observation that in molecules the correlation hole is not sufficiently accurate at medium range in RPA,  this has recently been extended to a non-local correction scheme  \cite{Ruzsinszky/etal:2010,Ruzsinszky/etal:2011}. Similarly, range-separated frameworks \cite{Toulouse/Colonna/Savin:2004} have been tried, in which only the long-range part of RPA is explicitly included \cite{Janesko/Henderson/Scuseria:2009, Janesko/Henderson/Scuseria:2009b,Toulouse/etal:2009,Zhu/etal:2010,Jansen/etal:2010,Toulouse/etal:2011,Angyan/etal:2011,Irelan/Henderson/Scuseria:2011}, whereas short/mid-range correlation is treated differently. Omitting the short-range part in RPA is also numerically beneficial whereby the slow convergence with respect to the number of basis functions can be circumvented. Due to this additional appealing fact, range-separated RPA is now an active research domain despite the empirical parameters that govern the range separation.  Another route to improve RPA in the framework of ACFD is to add an $f_\text{xc}$ kernel to the response function and to find suitable approximations for it \cite{Furche/Voorhis:2005,Hesselmann/Goerling:2010,Hesselmann/Goerling:2011,Hesselmann/Goerling:2011b}. Last but not least, the CCD perspective offers a different correction in form of the second-order screened exchange (SOSEX) contribution \cite{Freeman:1977,Grueneis/etal:2009,Paier/etal:2010}, whereas the MBPT perspective inspired single excitation (SE) corrections \cite{Ren/etal:2011}. SOSEX and SE are distinct many-body corrections and can also be combined \cite{Paier/etal:2012}. These corrections have a clear diagrammatic representation and alleviate the above-mentioned underbinding problem of standard RPA considerably \cite{Paier/etal:2012}. Yet another proposal to improve RPA by  incorporating higher-order exchange effects in various ways has also been discussed  recently \cite{Angyan/etal:2011}. However, at this point in time, a consensus regarding the ``optimal" correction that  combines both efficiency and accuracy has not been reached.

Although the majority of practical RPA calculations are performed as post-processing of a preceding DFT calculation, self-consistent RPA calculations  have also been performed within the optimized effective potential (OEP) framework. OEP is a procedure to find the optimal local multiplicative potential that minimizes orbital-dependent energy functionals. The first RPA-OEP calculations actually date back more than 20 years, but were not recognized as such. Godby, Schl\"uter and Sham solved the Sham-Schl\"uter equation for the $GW$ self-energy \cite{Godby/Schlueter/Sham:1986,Godby/Schluter/Sham:1988}, which  is equivalent to the RPA-OEP equation, for  the self-consistent RPA KS potential of bulk silicon and other semiconductors, but did  not calculate RPA ground-state energies. 
Similar calculations for other bulk materials followed later by Kotani \cite{Kotani:1998} and Gr\"uning {\it et al.}\cite{Gruening/Marini/Rubio:2006, Gruening/Marini/Rubio_2:2006}. Hellgren and von Barth  \cite{Hellgren/Barth:2007} and then later Verma and Bartlett \cite{Verma/Bartlett:2012} have looked at closed-shell atoms and observed that the OEP-RPA KS potential there reproduces the exact asymptotic behavior in the valence region, although its  behavior near the nucleus is not very accurate. Extensions to diatomic molecules have also appeared recently \cite{Hellgren/etal:2012,Verma/Bartlett:2012}. Our own work on the  SE correction to RPA indicates that the input-orbital dependence in RPA post-processing calculations is a significant issue. Some form of self-consistency would therefore be desirable. However, due to  the considerable numerical effort associated with OEP-RPA calculations, practical RPA calculations will probably remain of  the  post-processing type in the near future.

Despite RPA's appealing features its widespread use in chemistry and materials science is impeded by its computational cost, which is considerable compared to conventional (semi)local DFT functionals. Furche's original implementation based on a molecular particle-hole basis scales as $O(N^6)$ \cite{Furche:2001}. This can be reduced to $O(N^5)$ using the plasmon-pole formulation of RPA \cite{Furche:2008}, or to $O(N^4)$ \cite{Eshuis/Yarkony/Furche:2010} when the resolution-of-identity (RI) technique is employed. Scuseria \textit{et al.} \cite{Scuseria/Henderson/Sorensen:2008} pointed out even slightly earlier that a $O(N^4)$ scaling can be achieved by combining the ``ring-CCD" RPA formulation and the Cholesky decomposition technique. Our own RPA implementation \cite{Ren/etal:preprint} in FHI-aims \cite{Blum/etal:2009}, which has been used in production calculations \cite{Ren/Rinke/Scheffler:2009,Ren/etal:2011,Paier/etal:2012} before, is based on localized numeric atom-centered orbitals and the RI technique, and hence naturally scales as $O(N^4)$. Plane-wave based implementations \cite{Fuchs/Gonze:2002,Harl/Schimka/Kresse:2010} also automatically have $O(N^4)$ scaling.  However, in standard implementations the convergence with respect to unoccupied states is slow. Proposals to eliminate the dependence on the unoccupied states \cite{Nguyen/deGironcoli:2009,Wilson/Gygi/Galli:2008} in the context of plane wave bases, by obtaining the response function from  density-functional perturbation theory \cite{Baroni/etal:2001}, have not been explored so far for local-orbital based implementations. In local orbital based approaches the $O(N^4)$ scaling can certainly be reduced by exploiting matrix sparsity, as demonstrated recently in the context of $GW$  \cite{Foerster/etal:2011} or second-order M{\o}ller-Plesset perturbation theory (MP2) \cite{Ochsenfeld/etal:2007}. Also approximations to RPA \cite{Ismail-Beigi:2010} or effective screening models \cite{Nguyen/Galli:2010} might significantly improve the scaling and the computational efficiency. Recently, RPA has been cast into the continuum mechanics formulation of DFT \cite{Tao/etal:2009} with considerable success in terms of computational efficiency \cite{Gould/Dobson:2011}. In general, there is still room for improvement, which, together with the rapid increase in computer power makes us confident that RPA-type approaches will become a powerful technique in computational chemistry and materials science in the future. It would thus be desirable, if the material science community would start to build up benchmark sets for materials science akin to the ones in quantum chemistry (e.g. G2  \cite{Curtiss/etal:1991} or S22 \cite{Jurecka/etal:2006}). These should include prototypical bulk crystals, surfaces, and surface adsorbates and would aid the development of RPA-based approaches.

\section{\label{sec:theory}Theory and concepts}
	RPA can be formulated within different theoretical frameworks. One particularly convenient approach
to derive RPA is the so-called ``adiabatic connection (AC)",  which is a powerful mathematical 
technique to obtain the ground-state total energy of an interacting many-particle system. 
Starting with the AC approach, the interacting ground-state energy can be retrieved
either by coupling to the fluctuation-dissipation theorem in the DFT context, or by invoking
the Green-function based MBPT. RPA can be derived within both frameworks. In addition, 
RPA is also intimately linked to the coupled-cluster (CC) theory. In this section, we 
will present the theoretical aspects of RPA from several different perspectives.

\subsection{\label{sec:ac}Adiabatic connection}
The ground-state total energy of an interacting many-body Hamiltonian can formally be obtained
via the AC technique, in which a continuous set of coupling-strength ($\lambda$) dependent
Hamiltonians is introduced
\begin{equation}
 \hat{H}(\lambda) = \hat{H}_0 + \lambda \hat{H}_1(\lambda),
\label{Eq:H_lambda_1}
\end{equation}
that ``connect" a reference Hamiltonian $\hat{H}_0 = \hat{H}(\lambda=0)$ with the target many-body 
Hamiltonian $\hat{H}=\hat{H}(\lambda=1)$.
For the electronic systems considered here, $\hat{H}(\lambda)$ has the following form:
\begin{equation}
 \hat{H}(\lambda) 
   = \sum_{i=1}^N \left[ -\frac{1}{2}\nabla^2_i + v^\text{ext}_\lambda(i)\right] + 
     \sum_{i>j=1}^N \frac{\lambda}{|\bfr_i-\bfr_j|}, 
\label{Eq:H_lambda}
\end{equation}
where $N$ is the number of electrons, $v^\text{ext}_\lambda$ is a $\lambda$-dependent 
external potential with  $v_{\lambda=1}^\text{ext}(\bfr)=v^\text{ext}(\bfr)$ being 
the physical external potential of the fully-interacting system. Note that 
in general $v^\text{ext}_\lambda$ can be non-local in space for $\lambda \ne 1$. Hartree atomic 
units $\hbar=e=m_\text{e}=1$ are used here and in the following.
The reference Hamiltonian $H_0$, given by Eq.~\refeq{Eq:H_lambda} for $\lambda=0$, is of the
mean-field (MF) type, i.e., a simple summation over single-particle Hamiltonians:
\begin{eqnarray}
  \hat{H}_0 & = &\sum_{i=1}^N \left[ -\frac{1}{2}\nabla^2_i + v^\text{ext}_{\lambda=0} (i) \right] \nonumber \\
   &=& \sum_{i=1}^N \left[ -\frac{1}{2}\nabla^2_i + v^\text{ext}(\bfr_i) + v^\text{MF}(i) \right].
\label{Eq:H_ref}
\end{eqnarray}
In Eq.~\refeq{Eq:H_ref},  $v^\text{MF}$ is a certain (yet-to-be-specified) mean-field potential arising from the electron-electron interaction. It can be the Hartree-Fock (HF) potential $v^\text{HF}$ or the Hartree plus exchange-correlation potential $v^\text{Hxc}$ in DFT. Given Eq.~\eqref{Eq:H_lambda}
and \eqref{Eq:H_ref},  the perturbative Hamiltonian $\hat{H}_1(\lambda)$ in Eq.~\refeq{Eq:H_lambda_1} becomes
\begin{align}
&\hat{H}_1(\lambda)=\sum_{i>j=1}^N \frac{1}{|\bfr_i-\bfr_j|} + \frac{1}{\lambda} \sum_{i=1}^N 
 \left[ v^\text{ext}_\lambda(\bfr_i) - v^\text{ext}_{\lambda=0}(i) \right], \nonumber \\
&=\sum_{i>j=1}^N \frac{1}{|\bfr_i-\bfr_j|} + 
 \frac{1}{\lambda} \sum_{i=1}^N 
\left[ v^\text{ext}_\lambda(\bfr_i) - v^\text{ext} (\bfr_i) - v^\text{MF}(i) \right]\, .  \nonumber \\
\label{Eq:pert_ham}
\end{align}
In the AC construction of the total energy, we introduce the ground-state
wave function $|\Psi_\lambda \rangle$ for the $\lambda$-dependent system such that
\begin{equation}
 H(\lambda)|\Psi_\lambda \rangle = E(\lambda) |\Psi_\lambda \rangle\, .
\end{equation}
Adopting the normalization condition $\langle \Psi_\lambda | \Psi_\lambda \rangle =1$, 
the interacting ground-state total energy can then be obtained with the aid of the
Hellmann-Feynman theorem,
\begin{align}
E(\lambda=1) = & E_0 + \int_0^1 d\lambda \times \nonumber \\
    &  \langle \Psi_\lambda | \left(\hat{H}_1(\lambda) + \lambda 
        \frac{d \hat{H}_1(\lambda)}{d \lambda} \right)| \Psi_\lambda \rangle \, , 
\label{Eq:E_AC} 
\end{align}
where 
\begin{equation}
E_0 = E^\text{(0)} = \langle \Psi_0 | H_0 |\Psi_0\rangle
\end{equation}
is the zeroth-order energy.
We note that the choice of the adiabatic-connection path in Eq.~(\ref{Eq:E_AC}) is not unique. 
In DFT, the path is chosen such that the electron density is kept fixed at its physical value
along the way. This implies a non-trivial (not explicitly known) $\lambda$-dependence of 
$\hat{H}_1(\lambda)$. 
In MBPT, one often chooses a linear connection path --- $\hat{H}_1(\lambda)=\hat{H}_1$ 
(and hence  $d \hat{H}_1(\lambda)/ d \lambda=0$). In this case, a Taylor expansion of 
$|\Psi_\lambda\rangle$ in terms of $\lambda$ in Eq.~(\ref{Eq:E_AC}) leads to standard 
Rayleigh-Schr{\"o}dinger perturbation theory (RSPT) \cite{Szabo/Ostlund:1989}.

\subsection{\label{sec:acfd}RPA derived from ACFD}
Here we briefly introduce the concept of RPA in the context of DFT, which serves as the foundation for
most practical RPA calculations in recent years. In Kohn-Sham (KS) DFT, the ground-state total
energy for an interacting $N$-electron system is an (implicit) functional of the electron density $n(\bfr)$ and can be conveniently split into four terms: 
\begin{equation}
 E[n(\bfr)] = T_\text{s}[\psi_i(\bfr)] + E_\text{ext}[n(\bfr)] + E_\text{H}[n(\bfr)] + E_\text{xc}[\psi_i(\bfr)]\, .
\label{Eq:E_KS-DFT}
\end{equation}
$T_\text{s}$ is the kinetic energy of the KS independent-particle system, $E_\text{ext}$ the energy due to external 
potentials,
$E_\text{H}$ the classic Hartree energy, and $E_\text{xc}$ the exchange-correlation energy. In the KS framework,
the electron density is obtained from the single-particle KS orbitals $\psi_i(\bfr)$ via
$n(\bfr)=\sum_i^\text{occ} |\psi_i(\bfr)|^2$. Among the four terms in Eq.~\refeq{Eq:E_KS-DFT},
only $E_\text{ext}[n(\bfr)]$ and $E_\text{H}[n(\bfr)]$ are explicit functionals of $n(\bfr)$. $T_\text{s}$
is treated exactly in KS-DFT in terms of the single-particle orbitals $\psi_i(\bfr)$ which themselves are
functionals of $n(\bfr)$. 

All the many-body complexity is contained in the unknown XC energy term,
which is approximated as an explicit functional of $n(\bfr)$ (and its local gradients) in conventional functionals
(LDA and GGAs), and as a functional of the $\psi_i(\bfr)$'s in more advanced functionals (hybrid density
functionals, RPA, etc.). Different existing approximations to $E_\text{xc}$ can be classified into
a hierarchical scheme known as ``Jacob's ladder" \cite{Perdew/Schmidt:2001} in DFT.  However, what if one would like to improve the accuracy of $E_\text{xc}$ in a more systematic way?
For this purpose it is illuminating to start with the formally exact way of constructing $E_\text{xc}$
using the AC technique discussed above. As alluded to before, in KS-DFT the AC path is chosen
such that the electron density is kept fixed. 
Equation (\ref{Eq:E_AC}) for the exact ground-state total-energy $E=E(\lambda=1)$ then reduces to 
\begin{align}
 E= E_0  + & \int_0^1 d\lambda \langle 
    \Psi_\lambda | \frac{1}{2}\sum_{i\ne j=1}^{N}\frac{1}{|\bfr_i-\bfr_j|} |\Psi_\lambda \rangle \nonumber \\
    + & \int_0^1 d\lambda \langle\Psi_\lambda |\sum_{i=1}^N \frac{d}{d\lambda} v_\lambda^\text{ext}(\bfr_i)
        |\Psi_\lambda \rangle  \nonumber \\
    = E_0  + &\frac{1}{2}\int_0^1 d\lambda \iint d\bfr d\bfrp \times \nonumber \\
         &
         \langle \Psi_\lambda | \frac{ \hat{n}(\bfr)\left[\hat{n}(\bfrp) - 
         \delta(\bfr-\bfrp)\right] }{|\bfr-\bfrp|}
                  |\Psi_\lambda \rangle \nonumber \\
           + & \int d\bfr n(\bfr) \left[  v_{\lambda=1}^\text{ext}(\bfr) - 
             v_{\lambda=0}^\text{ext}(\bfr) \right] \, ,
\end{align}
where 
\begin{equation}
 \hat{n}(\bfr)=\sum_{i=1}^N \delta(\bfr -\bfr_i)
\end{equation}
is the electron-density operator, and
$n(\bfr) = \langle \Psi_\lambda | \hat{n}(\bfr) | \Psi_\lambda \rangle$ for any $0\le \lambda \le 1$.

For the KS reference state  $|\Psi_0\rangle$ (given by the Slater determinant of the occupied single-particle KS orbitals $\{\psi_i(\bfr)\}$) we obtain
  \begin{eqnarray}
     E_0 &=& \langle \Psi_0 | \sum_{i=1}^N \left[ -\frac{1}{2}\nabla^2 + v_{\lambda=0}^\text{ext}(\bfr_i) \right] |\Psi_0 \rangle
             \nonumber \\
         &=& T_\text{s}\left[\psi_i(\bfr)\right] + \int d\bfr n(\bfr) v^\text{ext}_{\lambda=0}(\bfr)\, ,
  \end{eqnarray}
and thus
 \begin{align}
  E = & T_\text{s}\left[\psi_i(\bfr)\right] + \int d\bfr n(\bfr) v_{\lambda=1}^\text{ext}(\bfr) + \nonumber \\
             &  \frac{1}{2}  \int_0^1 d\lambda \iint d\bfr d\bfrp 
         \langle \Psi_\lambda | \frac{ \hat{n}(\bfr)\left[\hat{n}(\bfrp) - 
         \delta(\bfr-\bfrp)\right] }{|\bfr-\bfrp|}
                  |\Psi_\lambda \rangle  \, .
  \label{Eq:E_AC_DFT}
 \end{align}
Equating \refeq{Eq:E_KS-DFT}  and \refeq{Eq:E_AC_DFT}, and noticing
 \begin{eqnarray}
   E_\text{H}[n(\bfr)]  &=& \frac{1}{2}\int d\bfr d\bfrp \frac{n(\bfr)n(\bfrp)}{|\bfr-\bfrp|} 
   \label{Eq:E_Hartree} \\
   E_\text{ext}[n(\bfr)] &=& \int d\bfr n(\bfr) v_{\lambda=1}^\text{ext}(\bfr)\, , 
 \end{eqnarray}
one obtains the formally exact expression for the XC energy
 \begin{equation}
   E_\text{xc} = \frac{1}{2} \int d\lambda \iint d\bfr d\bfrp 
   \frac{{n}^\lambda_\text{xc}(\bfr,\bfrp)n(\bfr)}{|\bfr-\bfrp|}\, .
  \label{Eq:AC}
 \end{equation}
Here 
\begin{equation}
 n_\text{xc}^\lambda(\bfr,\bfrp)=\frac{\langle \Psi_\lambda|\delta \hat{n}(\bfr)\delta\hat{n}(\bfrp)|\Psi_\lambda\rangle}
 {n(\bfr)} -\delta(\bfr-\bfrp)\, ,
 \label{Eq:XC_hole}
\end{equation}
is the formal expression for the so-called XC hole, with 
$\delta\hat{n}(\bfr)=\hat{n}(\bfr)-n(\bfr)$ being the 
\textit{fluctuation} of the density operator $\hat{n}(\bfr)$ around its expectation value $n(\bfr)$.
Equation \refeq{Eq:XC_hole} shows that the XC hole is related to the density-density correlation function. In physical terms,  it describes how the presence of an electron at point $\bfr$ depletes the density of all other electrons at another point $\bfrp$. 

In the second step, the density-density correlations (\textit{fluctuations}) in Eq.~\refeq{Eq:XC_hole} are linked to the response properties (\textit{dissipation}) of the system through the zero-temperature \textit{fluctuation-dissipation} theorem (FDT). The FDT is a powerful technique in statistical physics. It states that the response of a system at thermodynamic equilibrium to a small external perturbation is the same as its response to the spontaneous internal fluctuations in the absence of the perturbation \cite{Kubo:1966}.  The FDT is manifested in many physical phenomena and applies to both thermo and quantum-mechanical fluctuations.  The dielectric formulation of the many-body problem by  Nozi\'eres and Pines \cite{Nozieres/Pines:1958b} is a key example of the latter. In this context, the zero-temperature FDT leads to \cite{Nozieres/Pines:1966}
\begin{equation}
 \langle \Psi_\lambda|\delta\hat{n}(\bfr)\delta\hat{n}(\bfrp)|\Psi_\lambda\rangle = -
  \frac{1}{\pi}\int_{0}^\infty d\omega \text{Im} \chi^\lambda (\bfr, \bfrp, \omega)\, ,
 \label{Eq:FDT}
\end{equation}
where $\chi^\lambda (\bfr, \bfrp, \omega)$ is the linear density response function of the ($\lambda$-scaled) 
system.
Using Eqs.~(\ref{Eq:AC}-\ref{Eq:FDT}) and $v(\bfr,\bfrp) = 1/|\bfr-\bfrp|$, we arrive at the renowned ACFD expression for the XC energy in DFT
\begin{eqnarray}
E_\text{xc} &=& \frac{1}{2} \int_0^1 d\lambda \iint d\bfr d\bfrp v(\bfr,\bfrp) \times \nonumber \\ 
            && \left[ 
    -\frac{1}{\pi}\int_0^\infty d\omega \text{Im} \chi^\lambda (\bfr, \bfrp, \omega) - \delta(\bfr-\bfrp)n(\bfr) \right] \nonumber \\
       &=& \frac{1}{2\pi} \int_0^1 d\lambda \iint d\bfr d\bfrp  v(\bfr,\bfrp) \times \nonumber \\ 
       && \left[  -\frac{1}{\pi}\int_0^\infty d\omega 
        \chi^\lambda (\bfr, \bfrp, i\omega) - \delta(\bfr-\bfrp)n(\bfr) \right]\,  . \nonumber \\
 \label{Eq:EC-ACFD}
\end{eqnarray}
The reason that the above frequency integration can be performed along the imaginary axis originates
from the analytical structure of $\chi^\lambda (\bfr, \bfrp, \omega)$ and the fact that it becomes
real on the imaginary axis. The ACFD expression in Eq.~\refeq{Eq:EC-ACFD} 
transforms the problem of computing the XC energy to one of computing
the response functions of a series of fictitious systems along the AC path, 
which in practice have to be approximated as well. 

In this context the \textit{random-phase approximation} is a particularly simple approximation of the
response function:
\begin{eqnarray}
 &&  \chi^{\lambda}_{\text{RPA}}(\bfr, \bfrp, i\omega) = \chi^0(\bfr,\bfrp,i\omega) + \nonumber \\
  &&  \int d\bfr_1 d\bfr_2 \chi^0(\bfr,\bfr_1,i\omega)
   \lambda v(\bfr_1-\bfr_2)\chi^{\lambda}_{\text{RPA}}(\bfr_2,\bfrp,\omega) . \nonumber \\
 \label{Eq:RPA_response}
\end{eqnarray}
$\chi^0(\bfr,\bfr_1,i\omega)$ is the independent-particle response function of the 
KS reference system at $\lambda=0$ and  is known explicitly in terms of the single-particle KS orbitals $\psi_i(\bfr)$, orbital energies $\epsilon_i$ and occupation factors $f_i$
\begin{equation}
\chi^0(\bfr,\bfrp,i\omega)  = \sum_{ij}\frac{(f_i-f_j)\psi_i^\ast(\bfr)\psi_j(\bfr) \psi_j^\ast(\bfrp)\psi_i(\bfrp)}
 {\epsilon_i - \epsilon_j -i\omega}\, .
\label{Eq:indep_response}
\end{equation}
From equations (\ref{Eq:EC-ACFD}) and (\ref{Eq:RPA_response}), the XC energy in RPA can be separated 
into an exact exchange (EX) and the RPA correlation term,
\begin{equation}
E_\text{xc}^\text{RPA} = E_\text{x}^\text{EX} + E_\text{c}^\text{RPA},
\end{equation}
where
\begin{eqnarray}
 E_\text{x}^\text{EX}
       &=& \frac{1}{2} \iint d\bfr d\bfrp  v(\bfr,\bfrp)  \times \nonumber \\
       && \left[  -\frac{1}{\pi}\int_0^\infty d\omega 
        \chi^0 (\bfr, \bfrp, i\omega) - \delta(\bfr-\bfrp)n(\bfr) \right]\,   \nonumber \\
       &=& -\sum_{ij}f_if_j\iint d\bfr d\bfrp \psi_i^\ast(\bfr) \psi_j(\bfr) v(\bfr,\bfrp)
         \psi_j^\ast(\bfrp)\psi_i(\bfrp)\,  \label{Eq:E_EX} \nonumber \\
\end{eqnarray}
and
\begin{eqnarray}
E_\text{c}^\text{RPA}  &=& -\frac{1}{2\pi} \iint d\bfr d\bfrp  v(\bfr,\bfrp)  \times \nonumber \\
   && \int_0^\infty d\omega \left[ \int_0^1 d\lambda \chi^{\lambda}_{\text{RPA}}(\bfr, \bfrp, i\omega)
             -\chi_{0}(\bfr, \bfrp, i\omega) \right] \nonumber \\
   &=& \frac{1}{2\pi}\int_0^\infty d\omega \text{Tr} \left[ \text{ln}(1-\chi^0 (i\omega) v) + 
      \chi^0 (i\omega) v \right]. \nonumber \\
 \label{Eq:Ec_RPA}
\end{eqnarray}
For brevity the following convention 
 \begin{equation}
  \text{Tr}\left[AB\right] = \iint d\bfr d\bfrp A(\bfr,\bfrp)B(\bfrp,\bfr)
  \label{Eq:trace}
 \end{equation}
has been used in Eq. (\ref{Eq:Ec_RPA}).

\subsection{\label{sec:TH_MBPT}RPA derived from MBPT}
An alternative to ACFD is  to compute the interacting ground-state energy by performing an order-by-order expansion
of Eq.~(\ref{Eq:E_AC}). To this end, it is common to choose a linear AC path, i.e., 
in Eq.~\refeq{Eq:pert_ham} $v_\lambda^\text{ext} = v^\text{ext} + (1-\lambda)v^\text{MF}$  such that
 \begin{equation}
  \hat{H}_1(\lambda) = \hat{H}_1 = \sum_{i>j=1}^N \frac{1}{|\bfr_i-\bfr_j|}
 -\sum_{i=1}^N  v^\text{MF}_i\, .
 \end{equation}
Now equation \refeq{Eq:E_AC} reduces to
\begin{equation}
 E=E_0 + \int_0^1 d\lambda \langle \Psi_\lambda | \hat{H}_1 | \Psi_\lambda \rangle\, .
\end{equation}

A Taylor expansion of $|\Psi_\lambda \rangle$ and a subsequent $\lambda$ integration
lead to an order-by-order expansion of the interacting ground-state total energy,
e.g., the first-order correction to $E_0$ is given by
\begin{eqnarray}
E^\text{(1)} &=& \int_0^1 d\lambda \langle \Psi_0 | \hat{H}_1 |\Psi_0 \rangle \nonumber \\
             &=& \langle \Psi_0 | \hat{H}_1 |\Psi_0 \rangle \nonumber \\
             &=& E^\text{H} + E^\text{EX}_\text{x} - E^\text{MF}, 
\label{Eq:E_1st-order}
\end{eqnarray}
where $E^\text{H}$ and $E^\text{EX}_\text{x}$ are the classic Hartree and exact exchange energy defined in Eq.~\refeq{Eq:E_Hartree} and  \refeq{Eq:E_EX}, respectively. $E^\text{MF}= \langle \Psi_0 |v^\text{HF}|\Psi_0 \rangle$ is the ``double-counting" term due to the MF potential $v^\text{MF}$, which is already included in $H^0$. The sum of $E_0$ and the first-order term $E^\text{(1)}$ yields the Hartree-Fock energy, and all 
higher-order contributions constitute the so-called correlation energy.

\begin{figure}
  \begin{picture}(140,60)(0,0)
  \put(-40,20){$E_\text{c}^\text{MP2}=$}
  \put(0,0){\includegraphics[width=0.3\textwidth]{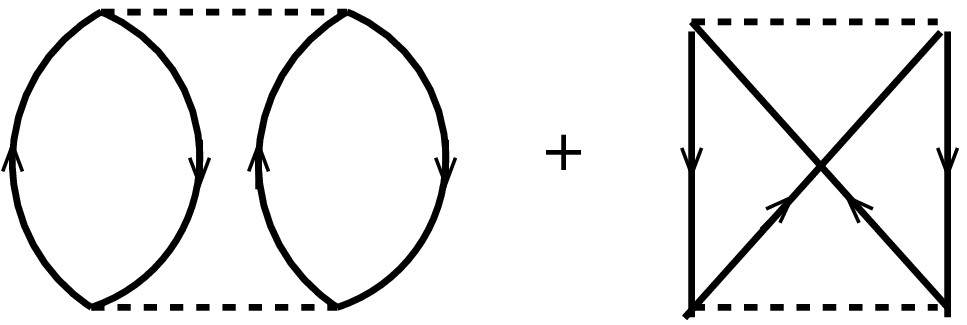}}
  \end{picture}
  \caption{\label{fig:MP2}Goldstone diagrams for the MP2 correlation energy. The two graphs describe respectively the
   second-order \textit{direct} process, and the second-order \textit{exchange} process.
   The upgoing solid line represents a particle associated with an unoccupied orbital energy $\epsilon_a$, 
   the downgoing solid line represents a hole associated with an occupied orbital energy $\epsilon_i$, 
   and the dashed line denotes the bare Coulomb interaction.} 
 \end{figure}

The higher-order terms can be evaluated using the diagrammatic technique developed by Goldstone  
\cite{Goldstone:1957}. 
For instance, the second-order energy in RSPT is given by
 \begin{align}
    E^{(2)}  = & \sum_{n>0}\frac{|\langle\Phi_0|\hat{H}_1|\Phi_n\rangle|^2}{E_0 - E^{(0)}_{n}} \nonumber \\
             = & \sum_{i}^\text{occ}\sum_{a}^\text{unocc}
               \frac{|\langle\Phi_0|\hat{H}_1|\Phi_{i}^{a}\rangle|^2}{E_0 - E^{(0)}_{i,a}} +
               \sum_{ij}^\text{occ}\sum_{ab}^\text{unocc}
               \frac{|\langle\Phi_0|\hat{H}_1|\Phi_{ij}^{ab}\rangle|^2}{E_0 - E^{(0)}_{ij,ab}} 
 \label{Eq:2PT}
 \end{align}
where $|\Phi_0\rangle=|\Psi_0\rangle$ is the ground state of the reference Hamiltonian $\hat{H}_0$, 
and $|\Phi_n\rangle$ for $n>0$
correspond to its excited states with energy $E^{(0)}_{n}=\langle \Phi_n| \hat{H}_0 | \Phi_n\rangle$.
$|\Phi_n\rangle$ can be classified into singly-excited configurations $|\Phi_{i,a}\rangle$, doubly-excited configurations
$|\Phi_{ij,ab}\rangle$, etc.. The summation in Eq.~(\ref{Eq:2PT}) terminates at the level of double excitations. This is
because $\hat{H}_1$ only contains one- and two-particle operators, and hence does not couple the ground state 
$|\Phi_0\rangle$ to triple and higher-order excitations. We will examine the single-excitation contribution in 
Eq.~\refeq{Eq:2PT} in detail in Section~\ref{sec:rSE_SOSEX}. Here it suffices to say that this term is zero for 
the HF reference and therefore is not included in MP2. The double-excitation contribution can be further
split into two terms, corresponding to the second-order \textit{direct} and \textit{exchange} energy in MP2, whose representation  in terms of Goldstone diagrams is depicted in Fig.~\ref{fig:MP2}. The rules to evaluate Goldstone diagrams can be
found in the classic book by Szabo and Ostlund \cite{Szabo/Ostlund:1989}. As a side remark, the application of MP2 used to be restricted to finite systems, and its extension to infinite periodic systems has been a big challenge because of 
its $O(N^5)$ canonical scaling.  In recent years, however, several authors demonstrated that it is in principle feasible to apply MP2 to one- or two-dimensional systems \cite{Sun/Bartlett:1996,Ayala/Kudin/Scuseria:2001,Hirata/etal:2004}.  With the more recent implementations in the CRYSCOR code \cite{Pisani/etal:2008} as well as in the VASP code \cite{Grueneis/Marsman/Kresse:2010}, the application of MP2 to three-dimensional crystalline solids has become realistic. 

The Goldstone approach is convenient for the lowest few orders, but becomes cumbersome or impossible for arbitrarily high orders, the evaluation of which is essential when an order-by-order perturbation breaks down and a ``\textit{selective summation to infinite order}" procedure has to be invoked. In this case, it is much  more convenient to express  the total energy in terms of the Green function and the self-energy, as done, e.g.,  
by Luttinger and Ward \cite{Luttinger/Ward:1961}. Using the Green function language,  
the ground-state total energy can be expressed as \cite{Luttinger/Ward:1961,Fetter/Walecka:1971},
\begin{align}
E& = E_0+\frac{1}{2}\int_0^1 \frac{d\lambda}{\lambda} \left(\frac{1}{2\pi} 
        \int_{-\infty}^\infty d\omega \text{Tr}\left[ G^0(i\omega) \Sigma(i\omega,\lambda)\right]\right)
     \label{Eq:MBPT_pert} \\
    &= E_0+ \frac{1}{2} \int_0^1 \frac{d\lambda}{\lambda} \left(\frac{1}{2\pi} 
      \int_{-\infty}^\infty d\omega \text{Tr}\left[G(i\omega,\lambda)\Sigma^\ast(i\omega,\lambda)
      \right]\right)
     \label{Eq:MBPT_sc}
\end{align}
where $G^0$ and $G(\lambda)$ are single-particle Green functions corresponding to the non-interacting
Hamiltonian $H_0$ and the scaled interacting Hamiltonian $H(\lambda)$, respectively. $\Sigma^{\ast}(\lambda)$ 
and $\Sigma(\lambda)$ are the proper (irreducible) and improper (reducible) self-energies of the interacting
system with interaction strength $\lambda$. [Proper self-energy diagrams are those which cannot be split into two 
by cutting a single Green function line.] Note that in Eq.~\eqref{Eq:MBPT_pert} and \eqref{Eq:MBPT_sc},  the trace convention of Eq.~\refeq{Eq:trace}  is implied. 

\begin{figure}
  \begin{picture}(400,210)(0,-50)
   \put(0,150){(a)}
   \put(0,115){ $\Sigma^{GW}_c(\lambda)=~\lambda^2$}
   \put(55,85){ \includegraphics{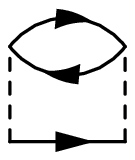}
    }
   \put(105,115){\bf $+~\lambda^3$}
   \put(117,85){\includegraphics{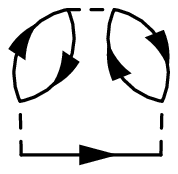}
    }
   \put(185,115){\bf $+~\cdots$}
   \put(0,80){(b)}
   \put(0,40){ $E_\text{c}^\text{RPA}=-~\frac{1}{4}$}
   \put(50,15){\includegraphics{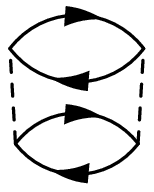}
    }
   \put(113,40){\bf $-~\frac{1}{6}$}
   \put(120,15){\includegraphics{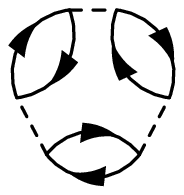}
    }
   \put(191,40){\bf $-~\cdots$}
   \put(0,10){(c)}
   \put(0,-25){ $E_\text{c}^\text{RPA}=$}
   \put(45,-40){ \includegraphics[width=0.34\textwidth]{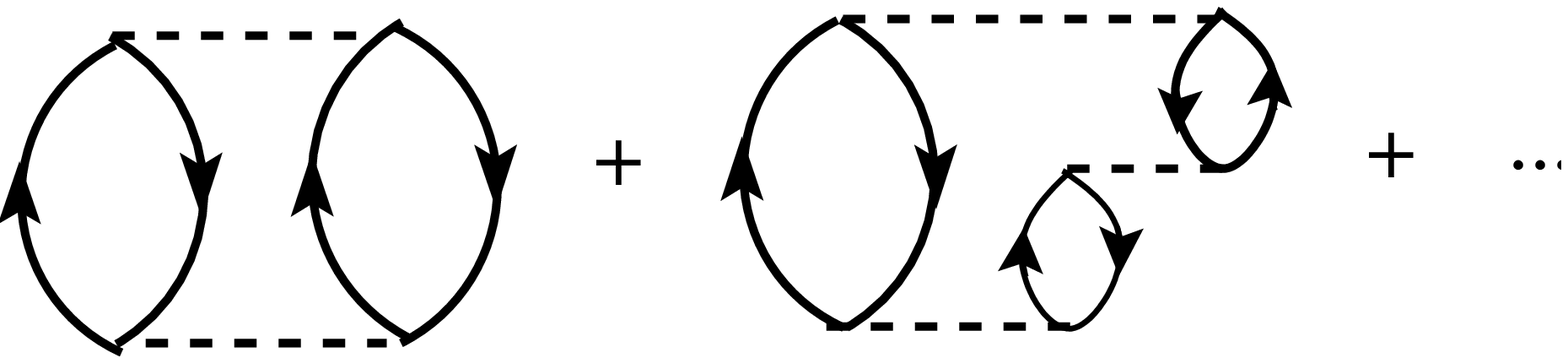}}
  \end{picture}
 \caption{\label{gw_diagram}Feynman diagrams for the $GW$ self-energy (a), Feynman diagrams for the RPA 
          correlation energy (b), and Goldstone diagrams for the RPA correlation energy (c).
          Solid lines in (a) and (b) (with thick arrows) represent fermion propagators $G$, and those in (c)
          (with thin arrows) denote particle (upgoing line) or hole states (down-going line) without frequency
          dependence. Dashed lines correspond to the bare Coulomb interaction $v$ in all graphs.}
 \end{figure}

The above quantities satisfy the following relationship
\begin{eqnarray}
G(i\omega,\lambda) &=& G^0(i\omega) + G^0(i\omega)\Sigma(i\omega,\lambda)G^0(i\omega) \nonumber \\
 &=& G^0(i\omega) + G^0(i\omega)\Sigma^\ast(i\omega,\lambda)G(i\omega,\lambda).
\label{Eq:Dyson}
\end{eqnarray}
From Eq.~(\ref{Eq:Dyson}) the equivalence of Eq.~(\ref{Eq:MBPT_pert}) and Eq.~(\ref{Eq:MBPT_sc}) is obvious.
In Eq.~\eqref{Eq:MBPT_pert}, a perturbation expansion of the $\lambda$-dependent self-energy
$\Sigma^\lambda(i\omega)$ naturally translates into a perturbation theory of the ground-state
energy. In particular, the linear term of $\Sigma^\lambda(i\omega)$ yields the first-order
correction to the ground-state energy, i.e., $E^{(1)}$ in Eq.~\eqref{Eq:E_1st-order}. 
All higher-order ($n\ge 2$) contributions of $\Sigma^\lambda(i\omega)$, here denoted
$\Sigma_\text{c}$, define the so-called correlation energy. In general the correlation energy cannot 
be treated exactly. A popular approximation to $\Sigma_\text{c}$ is the $GW$ approach,
which corresponds to a selective summation of self-energy diagrams with ring structure to infinite 
order, as illustrated in Fig.~\ref{gw_diagram}(a).
Multiplying~$G_0$ to the $GW$ self-energy 
$\Sigma_\text{c}^{GW}(i\omega)$ as done in Eq.~\eqref{Eq:MBPT_pert} and performing the $\lambda$ integration,
one obtains the RPA correlation energy 
\begin{equation}
  E_\text{c}^\text{RPA} = \frac{1}{2} \int_0^1 \frac{d\lambda}{\lambda} 
           \left(\frac{1}{2\pi}\int_{-\infty}^\infty d\omega 
           \text{Tr}\left[ G^0(i\omega) \Sigma_\text{c}^{GW}(i\omega,\lambda)\right]\right).
\end{equation}
This illustrates the close connection between RPA and the $GW$ approach.
A diagrammatic representation of $E_\text{c}^\text{RPA}$ is shown in Fig.~\ref{gw_diagram}(b).
We emphasize that the diagrams in Fig.~\ref{gw_diagram}(a) and \ref{gw_diagram}(b) are Feynman diagrams, i.e., the
arrowed lines should really be interpreted as propagators, or Green functions. A similar representation
of $E_\text{c}^\text{RPA}$ can be drawn in terms of Goldstone diagrams \cite{Szabo/Ostlund:1989}, as shown in
Fig~\ref{gw_diagram}(c).  However, caution should be applied, because the rules for evaluating these diagrams are 
different (see e.g., Ref. [\onlinecite{Fetter/Walecka:1971,Szabo/Ostlund:1989}]),
and the prefactors in Fig.~\ref{gw_diagram}(b) are not present in the corresponding Goldstone diagrams.
The leading term in RPA corresponds to the second-order \textit{direct} term in MP2.
\begin{fmffile}{gw}
\end{fmffile}

We note that starting from Eq.~\eqref{Eq:MBPT_pert} this procedure naturally gives the perturbative
RPA correlation energy based on any convenient non-interacting reference Hamiltonian $H_0$,
such as Hartree-Fock  or local/semi-local KS-DFT theory. If one instead starts with Eq.~\eqref{Eq:MBPT_sc} and applies the $GW$ approximation therein, $G(\lambda,i\omega)$ and $\Sigma^\ast(\lambda,i\omega)$ become the self-consistent
$GW$ Green function and self-energy. As a result  the improper self-energy diagrams in Eq.~\eqref{Eq:MBPT_pert}, which are neglected in the perturbative $GW$ approach (known as $G^0W^0$ in the literature), are introduced and the total energy differs from that of the RPA. An in-depth discussion of self-consistent $GW$ and its implications can be found in 
\cite{Holm/vonBarth:1998,Dahlen/Leeuwen/Barth:2005,Dahlen/Leeuwen/Barth:2006,Stan/Dahlen/Leeuwen:2006}.   

\subsection{\label{sec:rpa_ccd}Link to coupled cluster theory}
In recent years, RPA has also attracted considerable attention in the quantum chemistry community. 
One key reason for this is its intimate relationship with coupled cluster (CC) theory, which has been
very successful for accurately describing both covalent and non-covalent interactions in molecular systems. To understand
this relationship, we will give a very brief account of the CC theory here. More details
can for instance be found in a review paper by Bartlett and Musia{\l} \cite{Bartlett/Musial:2007}. The essential concept
of CC builds on the exponential ansatz for the many-body wave function $\Psi$ for correlated
electronic systems
\begin{equation}
  |\Psi\rangle = e^{\hat{T}}|\Phi\rangle .
 \label{Eq:CC_MBWF}
\end{equation}
$|\Phi\rangle$ is a non-interacting reference state, usually chosen to be the HF Slater determinant,
and $\hat{T}$ is a summation of excitation operators of different order,
 \begin{equation}
  \hat{T} = \hat{T_1} + \hat{T}_2 + \hat{T}_3 + \cdots +\hat{T}_n  + \cdots\, ,
  \label{Eq:T_expansion}
 \end{equation}
with $\hat{T}_1$, $\hat{T}_2$, $\hat{T}_3$, $\cdots$ being the single, double, and triple 
excitation operators, etc.
These operators can be most conveniently expressed using the language of second-quantization, namely,
\begin{eqnarray}
 \hat{T}_1 &=& \sum_{i,a}t_{i}^{a} \hat{c}^{\dagger}_a \hat{c}_i \nonumber, \\
 \hat{T}_2 &=& \frac{1}{4}\sum_{ij,ab}t_{ij}^{ab} \hat{c}^{\dagger}_a \hat{c}^{\dagger}_b \hat{c}_j \hat{c}_i,
  \label{Eq:double_amplitude} \\
    &\cdots& \nonumber \\
 \hat{T}_n &=& \frac{1}{(n!)^2}\sum_{ijk\cdots,abc\cdots}t_{ijk\cdots}^{abc\cdots} \hat{c}^{\dagger}_a \hat{c}^{\dagger}_b\hat{c}^{\dagger}_c\cdots \hat{c}_k \hat{c}_j \hat{c}_i,
\end{eqnarray}
where $\hat{c}^{\dagger}$ and $\hat{c}$ are single-particle creation and annihilation operators and $t_{i}^{a}$, $t_{ij}^{ab}$, $\ldots$ are the so-called CC singles, doubles, $\ldots$ amplitudes yet to be
determined. As before, $i,j,\ldots$ refer to occupied single-particle states, whereas $a,b,\cdots$ refer to 
unoccupied (\textit{virtual}) ones. Acting with $\hat{T}_n$ on the non-interacting
reference state $|\Phi_0\rangle$ generates $n$-order excited configuration denoted 
$|\Phi_{ijk\cdots}^{abc\cdots}\rangle$:
\begin{equation}
 \hat{T}_n |\Phi_0\rangle = \sum_{i>j>k\cdots,a>b>c\cdots}t_{ijk\cdots}^{abc\cdots} |\Phi_{ijk\cdots}^{abc\cdots}\rangle.
\end{equation}

The next question is how to determine the expansion coefficients $t_{ijk\cdots}^{abc\cdots}$? The CC many-body
wave function in Eq.~\refeq{Eq:CC_MBWF} has to satisfy the many-body Schr{\"o}dinger equation,
\begin{equation}
 \hat{H}e^{\hat{T}}|\Phi\rangle = Ee^{\hat{T}}|\Phi\rangle\, ,
\end{equation}
or 
\begin{equation}
e^{-\hat{T}} \hat{H}e^{\hat{T}}|\Phi\rangle = E|\Phi\rangle\, .
\label{Eq:CC_Schroedinger}
\end{equation}
By projecting Eq.~\refeq{Eq:CC_Schroedinger} onto the excited configurations $|\Phi_{ijk\cdots}^{abc\cdots}\rangle$, which have zero overlap with the non-interacting ground-state configuration $|\Phi\rangle$,
one obtains a set of coupled non-linear equations for the CC amplitudes $t_{ijk\cdots}^{abc\cdots}$, 
\begin{equation}
 \langle \Phi_{ijk\cdots}^{abc\cdots} | e^{-\hat{T}} \hat{H}e^{\hat{T}}|\Phi\rangle = 0\, .
 \label{Eq:CC_amplitude_equation}
\end{equation}
These can be determined by solving Eqs.~\refeq{Eq:CC_amplitude_equation} self-consistently. 

In analogy to the Goldstone diagrams, equation \refeq{Eq:CC_amplitude_equation} can be represented pictorially using diagrams, as illustrated by \v{C}\'{i}\v{z}ek \cite{Cizek:1966} in 1966. In practice, the expansion of the $\hat{T}$ operator has to be truncated. One popular choice is the CC doubles (CCD)  approximation, or $T_2$ approximation \cite{Cizek:1966}, that retains only the double excitation term in Eq.~\refeq{Eq:T_expansion}. The graphical representation of CCD contains a rich variety of diagrams including ring diagrams, ladder diagrams, the mixture of the two, etc. If one restricts the choice to the pure ring diagrams, as practiced in early work on the HEG \cite{Freeman:1977,Bishop/Luehrmann:1978}, 
the CCD equation is reduced to the following simplified form \cite{Scuseria/Henderson/Sorensen:2008},
\begin{equation}
  B + AT+ TA + TBT = 0 .
 \label{Eq:Riccati_equation}
\end{equation}
$A$, $B$, $T$ are all matrices of rank $N_\text{occ}\cdot N_\text{vir}$ with $N_\text{occ}$ and $N_\text{vir}$
being the number of occupied and unoccupied single-particle states, respectively. Specifically we have
$A_{ia,jb}= (\epsilon_i - \epsilon_a)\delta_{ij}\delta_{ab} -\langle ib|aj\rangle $, 
$B_{ia,jb} = \langle ij|ab\rangle$, and $T_{ia,jb}=t_{ij}^{ab}$, where the Dirac notation for the two-electron
Coulomb repulsion integrals 
\begin{equation}
 \langle pq|rs\rangle = \iint d\bfr d\bfrp \frac{\psi_p^\ast(\bfr)\psi_r(\bfr)\psi_q^\ast(\bfrp)\psi_s(\bfrp)}
  {|\bfr-\bfrp|} \, 
\end{equation}
has been adopted.

Equation (\ref{Eq:Riccati_equation}) is mathematically known as the Riccati equation \cite{Reid:1972}. 
Solving this equation yields the ring-CCD amplitudes $T^\text{rCCD}$, with
which the RPA correlation energy can be written as
\begin{equation}
   E_\text{c}^\text{RPA} = \frac{1}{2}\text{Tr}\left(BT^\text{rCCD}\right) = 
   \frac{1}{2}\sum_{ij,ab}B_{ia,jb}T_{jb,ia}^\text{rCCD}\, .
  \label{Eq:Ec_rpa_rCCD}
\end{equation}
The CCD formulation of RPA as given by Eq.~\refeq{Eq:Riccati_equation} and \refeq{Eq:Ec_rpa_rCCD} was   shown 
by Scuseria \textit{et al.} \cite{Scuseria/Henderson/Sorensen:2008} to be analytically equivalent to the plasmonic formulation of RPA. The latter has recently been discussed in detail by Furche \cite{Furche:2008,Eshuis/Bates/Furche:2011}, 
and hence will not be presented in this review. Technically, the solution of the Riccati equation 
\refeq{Eq:Riccati_equation} is not unique, due to the non-linear nature of the equation. One therefore has to make a judicious choice for the ring-CCD amplitudes in  practical RPA calculations \cite{Henderson/Scuseria:2010}. 


\section{\label{sec:comp}Algorithms and Implementations}
	\subsection{RPA implementations and scaling}
In this section we will briefly review different implementations of the RPA approach, since scaling and efficiency are particularly important for a computationally expensive approach like the RPA. Also, for historical reasons, the theoretical formulation of RPA is often linked closely to a certain implementation.  Similar to conventional DFT functionals, implementations of RPA 
can be based on local orbitals (LO), or on plane waves, or on (linearized) augmented plane waves (LAPW).
LO implementations have been reported for the development version of Gaussian  \cite{Janesko/Henderson/Scuseria:2009,Janesko/Henderson/Scuseria:2009b,Paier/etal:2010}, the development version of Molpro \cite{Toulouse/etal:2009,Hesselmann/Goerling:2010}, FHI-aims \cite{Ren/etal:preprint} and Turbomole \cite{Furche:2001,Eshuis/Yarkony/Furche:2010}. 
Plane-wave based implementations can be found in ABINIT \cite{Fuchs/Gonze:2002}, VASP  \cite{Harl/Kresse:2008,Harl/Schimka/Kresse:2010}, and  Quantum-Espresso \cite{Nguyen/deGironcoli:2009, Lu/Li/Rocca/Galli:2009}.  An early implementation by Miyake \textit{et al.}  \cite{Miyake/etal:2002} was based on LAPW. 

Furche's original implementation uses a molecular particle-hole basis and scales as $O(N^6)$ \cite{Furche:2001}, where $N$ is the number of atoms in the system (unit cell). This can be reduced to $O(N^5)$ using the plasmon-pole formulation of RPA \cite{Furche:2008}, or to $O(N^4)$ \cite{Eshuis/Yarkony/Furche:2010} when the resolution-of-identity (RI) technique is employed. Our own RPA implementation \cite{Ren/etal:preprint} in FHI-aims \cite{Blum/etal:2009} is described in Appendix \ref{sec:appendix_RI-FHI-aims}. It is based on localized numeric atom-centered orbitals and the RI technique, and hence naturally scales as $O(N^4)$. 

The key in the RI-RPA implementation is to expand the 
\textit{occupied}-\textit{virtual} orbital pair products 
$\phi_i^\ast(\bfr)\phi_j(\bfr)$ appearing in Eq.~\refeq{Eq:indep_response} in terms of a set of \textit{auxiliary} basis functions (ABFs) $\{P_{\mu}(\bfr)\}$. In this way, one can reduce the
rank of the matrix representation of $\chi^0$ from $N_\text{occ}*N_\text{vir}$ to $N_\text{aux}$ with $N_\text{aux}\ll N_\text{occ}*N_\text{vir}$. Here $N_\text{aux}$, $N_\text{occ}$, and $N_\text{vir}$ denote the number of
ABFs, and the numbers of occupied and unoccupied (\textit{virtual}) single-particle orbitals respectively.
With both $\chi^0$ and the coulomb kernel $v$ represented in terms of the ABFs, 
the RPA correlation energy expression in Eq.~\refeq{Eq:Ec_RPA} can be re-interpreted as a matrix equation of 
rank $N_\text{aux}$, which is numerically very cheap to evaluate. The dominating step then becomes the build 
of the matrix form
of $\chi^0$ which scales as $O(N^4)$. We refer the readers to Appendix~\ref{sec:appendix_RI-FHI-aims} and 
\refcite{Ren/etal:preprint} for further details.  

Plane-wave based implementations \cite{Fuchs/Gonze:2002,Harl/Schimka/Kresse:2010} automatically have $O(N^4)$ scaling. In a sense the plane-wave based RPA implementation is very similar in spirit
to the local-orbital-based RI-RPA implementation. In the former case the plane waves themselves serve as the
above-mentioned ABFs.

\subsection{Speed-up of RPA with iterative methods}
The RPA correlation energy in \refeq{Eq:Ec_RPA} can also be rewritten as follows,
\begin{equation}
  E_\text{c}^\text{RPA} = -\frac{1}{2\pi}\int_0^\infty d\omega \sum_{\mu}^{N_\text{aux}} \left[ \text{ln}
     \left(\varepsilon^\text{D}_\mu(i\omega)\right) 
     + 1 - \varepsilon^\text{D}_\mu(i\omega)\right], 
  \label{Eq:Ec_RPA_diag}
\end{equation}
where $\varepsilon^\text{D}_\mu(i\omega)$ is the $\mu$th eigenvalue of the dielectric function
$\varepsilon(i\omega)=1-\chi^0(i\omega)v$ represented in the ABFs. All eigenvalues
are larger than or equal to 1. From  \refeq{Eq:Ec_RPA_diag} it is clear that eigenvalues which are close to 1 have a vanishing contribution to the  correlation energy. For a set of different materials, Wilson \text{et al.} \cite{Wilson/Gygi/Galli:2008} observed that only a small fraction of the eigenvalues differs significantly from 1, which suggests that the full spectrum of $\varepsilon(i\omega)$ is not required for accurate RPA correlation energies. This opens up the possibility of computing the RPA correlation energy by obtaining the ``most significant" eigenvalues of $\varepsilon(i\omega)$ (or equivalently $\chi^0(i\omega)v$) from an iterative diagonalization procedure, instead of constructing and diagonalizing the full $\varepsilon(i\omega)$ or $\chi^0(i\omega)v$ matrices. 
In practice this can be conveniently done by resorting to the linear response technique of density functional perturbation theory (DFPT) \cite{Baroni/etal:2001} and has been proposed and implemented in the two independent works of Galli and coworkers  \cite{Wilson/Gygi/Galli:2008,Lu/Li/Rocca/Galli:2009}, and of Nguyen and de Gironcoli \cite{Nguyen/deGironcoli:2009} within a pseudopotential plane-wave framework. In these (plane-wave based) implementations the computational  cost is reduced from $N_{\text{pw-}\chi}^2N_\text{occ}N_\text{vir}$ to $N_{\text{pw-}\psi}N_\text{occ}^2N_\text{eig}$,  where $N_{\text{pw-}\chi}$ and $N_{\text{pw-}\psi}$ are the numbers of plane waves to expand the response function $\chi^0$ and the single-particle orbitals $\psi$ respectively, and $N_\text{eig}$ is the number of dominant eigenvalues. In this way, although the formal scaling is still $O(N^4)$, one achieves a large reduction of the prefactor, said to be 100-1000  \cite{Nguyen/deGironcoli:2009}. This procedure is in principle applicable to RI-RPA implementation in local-orbital basis sets as well, but has, to the best of our knowledge, not been reported so far.

\section{\label{sec:beyond}Computional schemes beyond RPA}
	In this Section we will give a brief account of the major activities for improving  the standard RPA, aiming at better accuracy.

\subsection{Semi-local and non-local corrections to RPA}
It is generally accepted that long-range interactions are well described within RPA,
whereas short-range correlations are not adequate \cite{Singwi/etal:1968}. This deficiency manifests itself most clearly in the pair-correlation function of the HEG, which spuriously becomes negative when the separation  between two electrons  gets small  \cite{Singwi/etal:1968,Hedin/Lundqvist:1969}. Based on this observation, Perdew and coworkers \cite{Kurth/Perdew:1999,Yan/Perdew/Kurth:2000} proposed a semi-local correction to RPA, termed as RPA+
 \begin{equation}
    E^\text{RPA+}_\text{c} = E_\text{c}^\text{RPA}  +  E_\text{c}^\text{GGA} -  E_\text{c}^\text{GGA-RPA},
   \label{Eq:Ec_RPA+}
 \end{equation}
where  $E_\text{c}^\text{GGA}$ is the GGA correlation energy, 
and $E_\text{c}^\text{GGA-RPA}$ represents the \textit{random-phase 
approximation} within GGA. 
Thus the difference between $E_\text{c}^\text{GGA}$ and $E_\text{c}^\text{GGA-RPA}$ gives
a semi-local correction to RPA for \textit{inhomogeneous} systems. As mentioned before
in the introduction, the RPA+ scheme, although conceptually appealing, and good for
total energies \cite{Jiang/Engel:2007}, does not significantly improve
 the description of \textit{energy differences}, in particular the atomization 
energies of small molecules \cite{Furche:2001}. This failure has been attributed to 
the inaccuracy of RPA in describing the multi-center non-locality of the correlation hole,
which cannot be corrected by semi-local corrections of the RPA+ type 
\cite{Ruzsinszky/etal:2010,Ruzsinszky/etal:2011}. A fully non-local correction (nlc) to RPA
has recently been proposed by Ruzsinszky, Perdew, and Csonka \cite{Ruzsinszky/etal:2011}. It takes  
the following form 
\begin{equation}
  E^\text{nlc}_\text{c} = \int d\bfr n(\bfr) \left[\epsilon^\text{GGA}(\bfr) - \epsilon^\text{GGA-RPA}(\bfr) \right]
     \left[ 1- \alpha F(f(\bfr)] \right],
  \label{Eq:Ec_RPA-nlc}
\end{equation}
where $\epsilon^\text{GGA}(\bfr)$ and $\epsilon^\text{GGA-RPA}(\bfr)$ are the GGA energy density
per electron and its approximate value within RPA, respectively.  $\alpha$ is an empirical parameter yet to be determined, 
and $F$ is a certain functional of $f(\bfr)$ -- the dimensionless ratio measuring the difference between the GGA exchange 
energy density and the exact-exchange energy density at a given point $\bfr$,
 \begin{equation}
    f(\bfr) = \frac{\epsilon_\text{x}^\text{GGA}(\bfr) - \epsilon_\text{x}^\text{exact}(\bfr)}
        {\epsilon_\text{x}^\text{GGA}(\bfr)}.
 \end{equation}
One may note that by setting $\alpha=0$ in Eq.~\refeq{Eq:Ec_RPA-nlc} the usual RPA+ correction term
is recovered. A simple choice of the functional form  $F(f)=f$ turns out to be good enough for fitting atomization energies, but the correct dissociation limit of H$_2$ given by the standard RPA is destroyed. To overcome this problem,
Ruzsinszky \textit{et al.} chose a more complex form of $F$,
\begin{equation}
 F(f) = f[1-7.2f^2][1+14.4f^2]\text{exp}(-7.2f^2),
\end{equation}
which ensures the correct dissociation limit, while yielding significantly improved 
atomization energies for $\alpha=9$. Up to now the correction scheme of Eq.~\refeq{Eq:Ec_RPA-nlc}
has not been widely benchmarked except for a small test set of 10 molecules where the atomization
energy has been improved by a factor of two \cite{Ruzsinszky/etal:2011}.

\subsection{Screened second-order exchange (SOSEX)}
\label{sec:sosex}
\begin{figure}
 \begin{picture}(200,60)(0,0)
  \put(-20,20){$E_\text{c}^\text{SOSEX}$}
  \put(20,0){\includegraphics[width=0.38\textwidth]{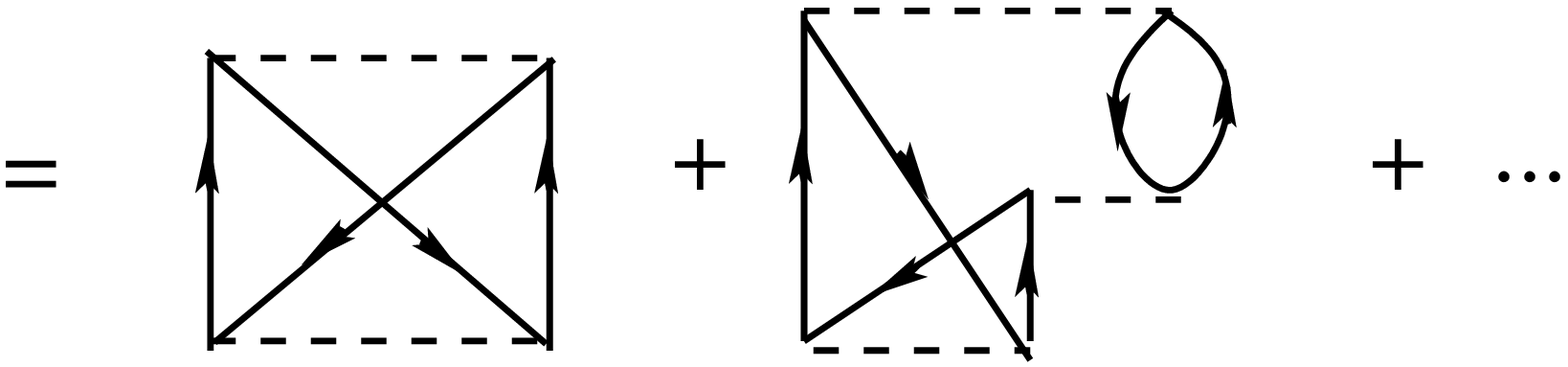}}
 \end{picture}
 \caption{\label{fig:Goldstone_SOSEX}Goldstone diagrams for SOSEX contribution. The rules
    to evaluate Goldstone diagrams can be found in \refcite{Szabo/Ostlund:1989}.}
\end{figure}

The SOSEX correction \cite{Freeman:1977,Grueneis/etal:2009,Paier/etal:2010} is an important route to go beyond  
standard RPA.  This concept can be most conveniently 
understood within the context of the ring-CCD formulation of RPA as discussed in \refsec{sec:rpa_ccd}.
If in Eq.~\refeq{Eq:Ec_rpa_rCCD},  the anti-symmetrized Coulomb integrals 
$\tilde{B}_{ia,jb}=\langle ij|ab\rangle - \langle ij|ba\rangle$ are inserted instead of the unsymmetrized Coulomb integrals,  the RPA+SOSEX correlation energy expression is obtained 
\begin{equation}
 E_\text{c}^\text{RPA+SOSEX} = \frac{1}{2}\sum_{ij,ab}T^\text{rCCD}_{ia,jb}\tilde{B}_{ia,jb}\, .
 \label{E_c_RPA+SOSEX}
\end{equation}
This approach, which was first used by Freeman \cite{Freeman:1977}, and recently examined by
Gr{\"u}neis \textit{et al.} for solids \cite{Grueneis/etal:2009} and Paier \cite{Paier/etal:2010}
for molecular properties, has received increasing attention in the RPA community. In contrast to RPA+, 
this scheme has the attractive feature that it improves both total energies and  energy differences simultaneously. 
Although originally conceived in the CC context, SOSEX has a clear representation in terms of Goldstone
diagrams, as shown in \Fref{fig:Goldstone_SOSEX} (see also \refcite{Grueneis/etal:2009}), which
can be compared to the Goldstone diagrams for RPA in Fig.~\ref{gw_diagram}(c).
From Fig.~\ref{fig:Goldstone_SOSEX}, it is clear that the leading term in SOSEX corresponds to the second-order
\textit{exchange} term of MP2.  In analogy, the leading term in RPA corresponds to the second-order
\textit{direct} term of MP2. Physically the second-order \textit{exchange} diagram describes a (virtual) process 
in which two particle-hole pairs are created spontaneously at a given time. The two particles 
(or equivalently the two holes) then exchange their positions, and these two (already exchanged) particle-hole pairs annihilate
themselves simultaneously at a later time. In SOSEX, similar to RPA, a sequence of higher-order diagrams are summed 
up to infinity. In these higher-order diagrams, after the initial creation and exchange process, one particle-hole
pair is scattered into new positions repeatedly following the same process as in RPA, until 
it annihilates simultaneously with the other pair at the end of the process. 

SOSEX is one-electron self-correlation free  and ameliorates the short-range over-correlation problem of RPA  to a large extent, leading to significantly better total energies \cite{Freeman:1977,Grueneis/etal:2009}. More importantly,  the RPA underestimation of atomization energies is substantially reduced. However, the dissociation of covalent diatomic molecules, which is well described in RPA, worsens considerably as demonstrated in \refcite{Paier/etal:2010} and to be shown in Fig.~\ref{Fig:Mol_Disso_asymp}.  It was argued that the self-correlation error present in RPA mimics static correlation, which becomes dominant in the dissociation limit of covalent molecules \cite{Henderson/Scuseria:2010}.

\subsection{Single excitation correction and its combination with SOSEX}
\label{sec:rSE_SOSEX}

In most practical calculations, RPA and SOSEX correlation energies are evaluated using
input orbitals from a preceding KS or generalized KS (gKS) \cite{GDFT:1993} calculation.
In this way both RPA and SOSEX can be interpreted as \textit{infinite-order summations of selected types of diagrams}
within the MBPT framework introduced in \refsec{sec:TH_MBPT}, as is evident from Figs~\ref{gw_diagram}(c) and
(\ref{fig:Goldstone_SOSEX}).
This viewpoint is helpful for identifying contributions missed in RPA through the aid of diagrammatic techniques.  An an example, the second-order energy in RSPT in Eq.~\refeq{Eq:2PT} have contributions from single
excitations (SE) and double excitations. The latter gives rise to the familiar MP2 correlation energy, which
is included in the RPA+SOSEX scheme as the leading term.  
The remaining SE term is given by
  \begin{align}
      E_\text{c}^\text{SE} = & \sum_{i}^\text{occ}\sum_{a}^\text{unocc}
               \frac{|\langle\Phi_0|\hat{H}_1|\Phi_{i}^{a}\rangle|^2}{E_0 - E^{(0)}_{i,a}} \nonumber \\
          =  & \sum_{ia}\frac{|\langle \psi_i | \hat{v}^\text{HF} - \hat{v}^\text{MF}|\psi_a \rangle |^2}
                  {\epsilon_i - \epsilon_a} \label{Eq:Ec_SE_deltaV} \\
          =  & \sum_{ia}\frac{|\langle \psi_i | \hat{f} |\psi_a \rangle |^2}
                  {\epsilon_i - \epsilon_a} 
     \label{Eq:Ec_SE}
  \end{align}
where $\hat{v}^\text{HF}$ is the self-consistent HF single-particle potential, $\hat{v}_\text{MF}$ is
the mean-field potential associated with the reference Hamiltonian, and 
$\hat{f}=-\nabla^2/2+\hat{v}^\text{ext} + \hat{v}^\text{HF}$
is the single-particle HF Hamiltonian (also known as the Fock operator in the quantum chemistry literature). 
A detailed derivation of Eq.~\eqref{Eq:Ec_SE_deltaV} using second-quantization  can be found in the 
supplemental material of Ref.~[\onlinecite{Ren/etal:2011}].  The equivalence of Eqs.~\refeq{Eq:Ec_SE} and 
\refeq{Eq:Ec_SE_deltaV} can be readily
confirmed by observing the relation between $\hat{f}$ and the single-particle reference
Hamiltonian $\hat{h}^\text{MF}$: $\hat{f}=\hat{h}^\text{MF} + \hat{v}^\text{HF}- \hat{v}^\text{MF}$, and
the fact $\langle \psi_i|\hat{h}^\text{MF}|\psi_a \rangle = 0$. Obviously for a HF reference 
where $\hat{v}^\text{MF}=\hat{v}^\text{HF}$, 
Eq.~\refeq{Eq:Ec_SE} becomes zero, a fact known as Brillouin theorem \cite{Szabo/Ostlund:1989}. 
Therefore, as mentioned in Section~\ref{sec:TH_MBPT}, this term is not present in MP2 theory
which is based on the HF reference.  
We note that a similar SE term also appears in 2nd-order
G\"orling-Levy perturbation theory (GL2) \cite{Goerling/Levy:1993,Jiang/Engel:2006}, \textit{ab initio} DFT 
\cite{Bartlett:2010}, as well as in CC theory \cite{Bartlett/Musial:2007}. However, the SE terms 
in different theoretical frameworks differ quantitatively. For instance, in GL2 $v^\text{MF}$ should be 
the exact-exchange OEP potential instead of the reference mean-field potential.

In \refcite{Ren/etal:2011} we have shown that adding the SE term of Eq.~(\ref{Eq:Ec_SE}) to
RPA significantly improves the accuracy of vdW-bonded molecules, which the standard RPA
scheme generally underbinds. This improvement carries over to
atomization energies of covalent molecules and insulating solids as shown in Ref.~[\onlinecite{Paier/etal:2012}].
It was also observed in \refcite{Ren/etal:2011} that a similar improvement can be achieved by replacing 
the non-self-consistent HF part of the RPA total energy by its self-consistent counterpart. It appears
that, by iterating the exchange-only part towards self-consistency, the SE effect can be accounted for 
effectively.  This procedure is termed ``hybrid-RPA", and has been
shown to be promising even for surface adsorption problems \cite{Goltl/Hafner:2011}.

\begin{figure}
 \begin{picture}(140,60)(0,0)
  \put(-40,20){$E_\text{c}^\text{rSE}=$}
  \put(0,0){\includegraphics[width=0.38\textwidth]{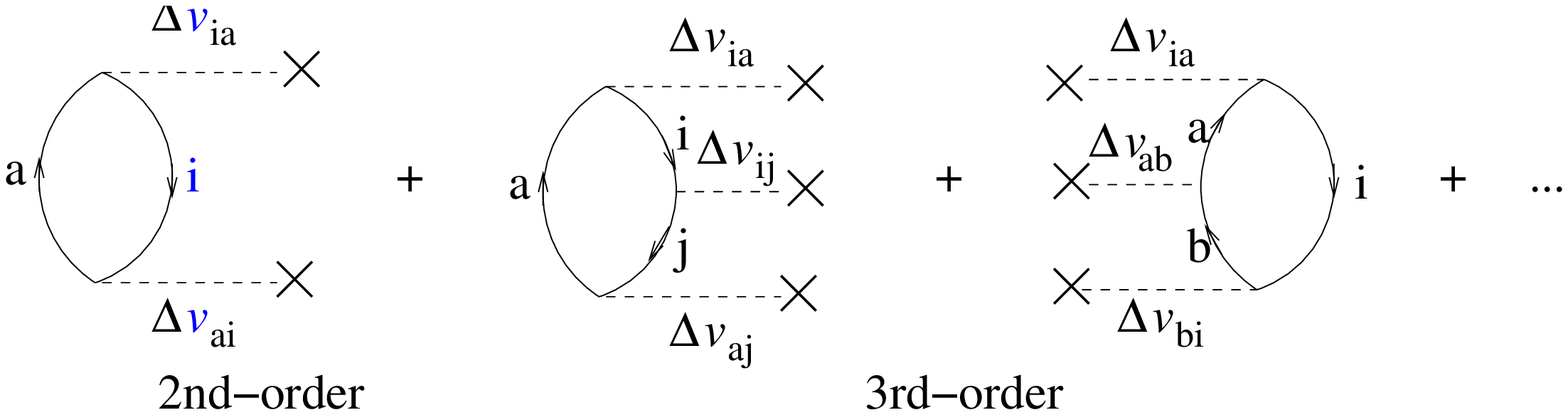}}
 \end{picture}
  \caption{\label{fig:rse_Goldstone}Goldstone diagrams for renormalized single excitation contributions. 
    Dashed lines ending with a cross denote the matrix element 
    $\Delta v_{pq} = \langle \psi_p| \hat{v}^\text{HF} - \hat{v}^\text{MF} |\psi_q\rangle$. }
\end{figure}

The SE energy in Eq.~\refeq{Eq:Ec_SE} is a second-order term in RSPT, which suffers from the same divergence problem as MP2 for systems with zero (direct) gap. To overcome this problem, 
in \refcite{Ren/etal:2011} we have proposed to sum over a sequence of higher-order diagrams involving only 
single excitations. This procedure can be illustrated in terms of Goldstone diagrams as shown in Fig.~\ref{fig:rse_Goldstone}.
 This summation follows the spirit of RPA and we denote it \textit{renormalized single 
excitations} (rSE) \cite{Ren/etal:2011}. The SE contribution to the 2nd-order correlation energy in Eq.~(\ref{Eq:Ec_SE}), represented by the first diagram in Fig~\ref{fig:rse_Goldstone}, constitutes the leading term in the rSE series.  A preliminary version of rSE, which neglects the ``off-diagonal" terms of the higher-order SE diagrams (by setting $i=j=\cdots$ 
and $a=b=\cdots$), was benchmarked for atomization energies and reaction barrier heights in 
\refcite{Paier/etal:2012}. Recently we were able to also include the ``off-diagonal" terms,
leading to a refined version of rSE.  
This rSE ``upgrade" does not affect the energetics of strongly bound molecules, as those benchmarked
in  Ref.~\cite{Paier/etal:2012}. However, the interaction energies of weakly bound molecules improve considerably.  A more detailed description of the computational procedure and extended benchmarks for rSE will be
reported in a forthcoming paper \cite{Ren/etal:inpreparation}.  However, we note that all the rSE results reported
in Section (\refsec{sec:appl}) correspond to the upgraded rSE.

Diagrammatically, RPA, SOSEX and rSE are three distinct infinite series of many-body terms, in which the three leading
terms correspond to the three terms in 2nd-order RSPT.  Thus it is quite natural to 
include all three of them, and the resultant RPA+SOSEX+rSE scheme can be viewed a \textit{renormalization} of 
the normal 2nd-order RSPT. Therefore we will refer to RPA+SOSEX+rSE as  ``\textit{renormalized second-order perturbation theory}" or r2PT in the following.

\subsection{Other ``beyond-RPA" activities}

There have been several other attempts to go beyond RPA. Here we will only briefly discuss the essential concepts behind these approaches without going into details. The interested reader is referred to the corresponding references. Following the ACFD formalism, as reviewed in \refsec{sec:acfd}, one possible route is to improve the interacting density response function. This can be conveniently done by adding the exchange correlation  kernel ($f_\text{xc}$) of time-dependent DFT 
\cite{Runge/Gross/:1984,Petersilka/Gossmann/Gross:1996}, that is omitted in RPA. Fuchs \textit{et al.} \cite{Fuchs/Gonze/Burke:2005},  as well as He{\ss}elmann and G{\"o}rling \cite{Hesselmann/Goerling:2011b} have added the exact-exchange kernel to RPA, a scheme termed by these authors as RPA+$X$ or EXX-RPA, for studying the H$_2$ dissociation problem. 
RPA+$X$ or EXX-RPA displays a similar dissociation behavior for H$_2$ as RPA: accurate at infinite separation, but 
slightly repulsive at intermediate bond lengths. This scheme however gives rise to a noticeable improvement of the total 
energy \cite{Hesselmann/Goerling:2010}.
Furche and Voorhis 
examined the influence of several different local and non-local kernels on the atomization energies of
small molecules and the binding energy curves of rare-gas dimers \cite{Furche/Voorhis:2005}. They found that semilocal
$f_\text{xc}$ kernels lead to a diverging pair density at small inter-particle distances, and it is necessary to
go to non-local $f_\text{xc}$ kernels to cure this. 
More work along these lines has to be done before conclusions can be drawn and accurate kernels become available.

Quantum chemistry offers another route to go beyond the standard RPA by including higher-order exchange effects (often termed as ``RPAx"). There the two-electron Coulomb integrals usually appear in an antisymmetrized form, whereby exchange-type contributions, which are neglected in standard RPA, are included automatically \cite{McLachlan/Ball:1964,Szabo/Ostlund:1977}. The RPA correlation energy can be expressed as a contraction between the ring-CCD amplitudes and the Coulomb integrals (see Eq.~\refeq{Eq:Ec_rpa_rCCD}), or alternatively between the coupling-strength-averaged density matrix and the Coulomb integrals (see \refcite{Jansen/etal:2010}). Different flavors of RPA can therefore be constructed depending on whether one antisymmetrizes the averaged density matrix and/or the Coulomb integrals (see {\'A}ngy{\'a}n \textit{et al.} \cite{Angyan/etal:2011}). According to our definitions in this article, these schemes are categorized as different ways to go beyond the standard RPA, while in the quantum chemistry community they might be simply referred to as RPA. The SOSEX correction, discussed  in \refsec{sec:sosex}, can also be rewritten in terms of a coupling-strength-averaged density matrix \cite{Jansen/etal:2010}.
Another interesting scheme was proposed by He{\ss}elmann \cite{Hesselmann:2011}, in which RPA is corrected to be exact at the third order of perturbation theory. These corrections show promising potential for the small molecules considered in \refcite{Hesselmann:2011}. However, more benchmarks are needed for a better assessment.

Both standard RPA and RPA with the exchange-type corrections discussed above have been tested in a range-separation framework by several authors \cite{Toulouse/etal:2009,Zhu/etal:2010,Janesko/Henderson/Scuseria:2009,Janesko/Henderson/Scuseria:2009b,Jansen/etal:2010,Toulouse/etal:2011,Angyan/etal:2011,Irelan/Henderson/Scuseria:2011}. As mentioned briefly in the introduction, the concept of range separation is similar to the RPA+ procedure, in which only the long-range behavior of RPA is retained. However, instead of additional corrections, here RPA at the short-range is completely removed and  replaced by semi-local or hybrid functionals. The price to pay is an empirical parameter that controls the range separation. The gain is better accuracy in describing molecular binding energies \cite{Janesko/Henderson/Scuseria:2009,Toulouse/etal:2009}, and increased computational efficiency. The latter is due to a reduction in the number of required basis functions to converge the long-range RPA part, which is no longer affected by the cusp condition. More details on range-separated RPA can be found in the original references \cite{Janesko/Henderson/Scuseria:2009,Janesko/Henderson/Scuseria:2009b,Toulouse/etal:2009,Zhu/etal:2010, Jansen/etal:2010,Toulouse/etal:2011,Angyan/etal:2011,Irelan/Henderson/Scuseria:2011}.
Compared to the diagrammatic approaches discussed before, the range-separation framework offers an 
alternative and computationally more efficient way to handle short-range correlations, albeit at the 
price of introducing some empiricism into the theory.

\section{\label{sec:appl}Applications}
	\subsection{\label{sec:mol}Molecules}
		RPA based approaches have been extensively benchmarked for molecular systems, ranging from 
the dissociation behavior of diatomic molecules \cite{Furche:2001,Fuchs/Gonze:2002,Janesko/Henderson/Scuseria:2009,Toulouse/etal:2009,Nguyen/Galli:2010,Irelan/Henderson/Scuseria:2011,Hesselmann/Goerling:2011b}, 
atomization energies of small covalent molecules \cite{Furche:2001,Ruzsinszky/etal:2010,Paier/etal:2010,Ruzsinszky/etal:2011,Paier/etal:2012}, interaction energies of weakly bonded molecular complex \cite{Zhu/etal:2010,Jansen/etal:2010,Toulouse/etal:2011,Angyan/etal:2011,Eshuis/Furche:2011}, and chemical reaction barrier heights \cite{Paier/etal:2012,Eshuis/Bates/Furche:2011}. The behavior of RPA for breaking covalent bonds was examined in its early days \cite{Furche:2001,Fuchs/Gonze:2002}, and today is still a topic of immense interest \cite{Toulouse/etal:2009,Nguyen/Galli:2010,Irelan/Henderson/Scuseria:2011,Hesselmann/Goerling:2011b}. Atomization energies of covalent molecules are somewhat disappointing, because standard RPA, as well as its local correction (RPA+), is not better than semi-local DFT functionals \cite{Furche:2001}. This issue was subsequently referred to as ``the RPA atomization energy puzzle" \cite{Ruzsinszky/etal:2010}. A solution can be found in the beyond-RPA schemes such as RPA+SOSEX \cite{Grueneis/etal:2009,Paier/etal:2010} and RPA+SE \cite{Ren/etal:2011,Paier/etal:2012,Ren/etal:inpreparation}.  Another major application area of RPA are weakly bonded molecules. Due to the seamless inclusion of the ubiquitous vdW interactions, RPA clearly improves over conventional DFT functionals, including hybrids. This feature is very important for systems where middle-ranged non-local electron correlations play a significant role, posing great challenges to empirical or semi-empirical pairwise-based correction schemes. Finally, for activation energies it turned out that standard RPA performs remarkably well \cite{Paier/etal:2012,Eshuis/Bates/Furche:2011} and the beyond RPA  correction schemes that have been developed so far do not improve the accuracy of the standard RPA \cite{Paier/etal:2012}. 

In the following, we will discuss the performance of RPA and its variants using representative examples to illustrate the aforementioned points. 

\subsubsection{\label{sec:Mol_Dissociation}Dissociation of diatomic molecules }
\begin{figure*}
 \includegraphics[width=0.6\textwidth,clip=true]{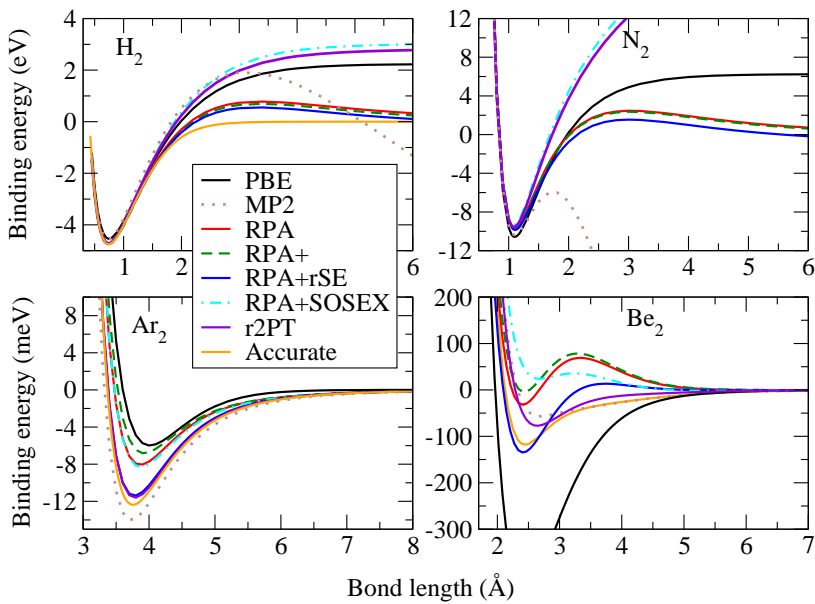}
 \caption{\label{Fig:Mol_Disso}Dissociation curves for H$_2$, N$_2$, Ar$_2$, and Be$_2$ using PBE, MP2, and 
    RPA-based methods.  All RPA-based methods use PBE orbitals as input. ``Accurate" reference curves
      are obtained with the full CI method for H$_2$ \cite{Wolniewicz:1993}, the Tang-Toennies potential model for Ar$_2$ 
     \cite{Tang/Toennies:2003}, and the extended germinal model for Be$_2$ \cite{Roeggen/Veseth:2005}. }
\end{figure*}

\begin{figure*}
 \includegraphics[width=0.6\textwidth,clip=true]{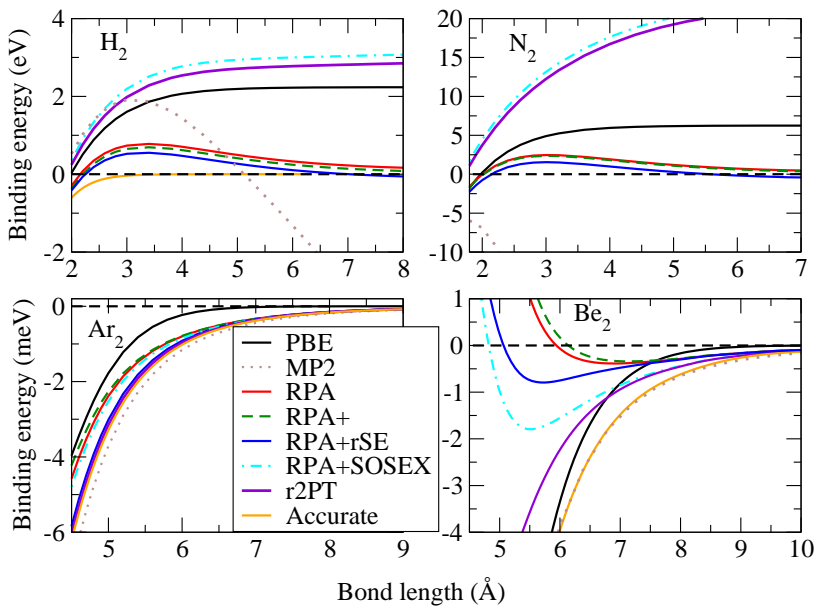}
 \caption{\label{Fig:Mol_Disso_asymp} Asymptotic region of the curves in  \Fref{Fig:Mol_Disso}.}
\end{figure*}

The dissociation of diatomic molecules is an important test ground for electronic structure methods. The performance of RPA on prototypical molecules has been examined in a number of studies \cite{Furche:2001,Fuchs/Gonze:2002,Janesko/Henderson/Scuseria:2009,Mori-Sanchez/etal:2012,Toulouse/etal:2009,Nguyen/Galli:2010,Irelan/Henderson/Scuseria:2011,Hesselmann/Goerling:2011b}. Here we present a brief summary of the behavior of RPA-based approaches based on data produced using our in-house code FHI-aims 
\cite{Blum/etal:2009}. The numerical details and benchmark studies of our RPA implementation have been
presented in \refcite{Ren/etal:preprint}.  In \Fref{Fig:Mol_Disso} the binding energy curves
obtained with PBE, MP2, and RPA-based methods are plotted for four molecular dimers, including two covalent molecules 
 (H$_2$ and N$_2$), one purely vdW-bonded molecule (Ar$_2$), and one with mixed character (Be$_2$).
Dunning's Gaussian cc-pV6Z basis \cite{Dunning:1989,Wilson/vanMourik/Dunning:1996} was used for H$_2$ and N$_2$, aug-cc-pV6Z for Ar$_2$, and cc-pV5Z for Be$_2$. Currently,  no larger basis  seems to be available for Be$_2$, but this will not affect the discussion here. Basis-set superposition errors (BSSE) are corrected using the
Boys-Bernardi counterpoise procedure \cite{Boys/Bernardi:1970}.
Also plotted in \Fref{Fig:Mol_Disso} are accurate theoretical reference data for H$_2$, Ar$_2$, and Be$_2$ coming respectively from the full CI approach \cite{Wolniewicz:1993}, the Tang-Toennies model \cite{Tang/Toennies:2003}, and the extended germinal model \cite{Roeggen/Veseth:2005}.
To visualize the corresponding asymptotic behavior more clearly, the large bond distance regime of all curves is 
shown in \Fref{Fig:Mol_Disso_asymp}.

RPA and RPA+ dissociate the covalent molecules correctly to their atomic limit at large separations, albeit from above after going through a positive ``bump" at intermediate bond distances. The fact that spin-restricted RPA calculations yield the correct H$_2$ dissociation limit is quite remarkable, given the fact that most spin-restricted single-reference methods, including local and semi-local DFT, Hartree-Fock, as well as the coupled cluster methods, yield an dissociation limit that is often too high in energy, as illustrated in \Fref{Fig:Mol_Disso} for PBE. MP2 fails more drastically, 
yielding diverging results in the dissociation limit for H$_2$ and N$_2$. The RPA+ binding curves follow the RPA ones closely, with only minor differences. The rSE corrections are also quite small in this case, shifting  the RPA curves towards larger binding energies, with the consequence that the binding energy dips slightly below zero in the dissociation limit (see the N$_2$ example in \Fref{Fig:Mol_Disso_asymp}).  This shift however leads to better molecular binding energies around the equilibrium where RPA systematically underbinds. The SOSEX correction, on the other hand, leads to dramatic changes. Although ``bump'' free, RPA+SOSEX yields dissociation limits that are much too large, even larger than PBE.  This effect carries over to r2PT at large bond distances  where rSE does not reduce the SOSEX overestimation.  

For the purely dispersion-bonded dimer Ar$_2$, all RPA-based approaches, as well as MP2, yield the correct $C_6/R^6$ asymptotic behavior, whereas the semi-local PBE functional gives a too fast exponential decay. Quantitatively, the $C_6$ 
dispersion coefficient is underestimated by $\sim 9\%$ within RPA (based on a PBE reference) \cite{Ren/unpublished}, 
and SOSEX or rSE will not change this. In contrast, MP2 overestimates the $C_6$ value by $\sim 18\%$ 
 \cite{Ren/unpublished}.  Around the equilibrium point, RPA and RPA+ underbind Ar$_2$ significantly. 
The rSE correction improves the results considerably, bringing  
the binding energy curve into close agreement with the Tang-Toennies reference curve. The SOSEX correction, on the
other hand, does very little in this case. As a consequence, r2PT resembles RPA+rSE closely in striking contrast 
to the covalent molecules.

Be$_2$ represents a more complex situation, in which both static correlation and long-range vdW interactions play an important role. In the intermediate regime, the RPA and RPA+ binding energy curves display a positive bump which is much more pronounced than for purely covalent molecules. At very large bonding distances, the curves cross the energy-zero line and eventually approach the atomic limit from below. The rSE correction moves the binding energy curve significantly further down, giving binding energies in good agreement with the reference values, whereas in the intermediate region a small positive bump remains. The SOSEX correction exhibits a complex behavior. While reducing the  bump it concomitantly weakens binding around the equilibrium distance. Combining the corrections from SOSEX and rSE, r2PT does well at intermediate and large bonding distances, but the binding energy at equilibrium is still noticeably too small. 
Regarding MP2, it is impressive to observe that this approach yields a binding energy curve that is in almost 
perfect agreement with the reference in the asymptotic region, although a substantial underbinding can be seen around 
the equilibrium region.

Summarizing this part, RPA with and without corrections shows potential, but at this point, none of
the RPA-based approaches discussed above can produce quantitatively accurate binding energy curves for all bonding situations. It is possible, but we consider it unlikely, that iterating RPA to self-consistency will change this result.
Apart from  applications to neutral molecules, RPA and RPA+SOSEX studies have been carried out for the dissociation of charged molecules. RPA fails drastically in this case 
\cite{Mori-Sanchez/etal:2012}, giving too low a total energy in the dissociation limit. Adding SOSEX
to RPA fixes this problem, although the correction now overshoots (with the exception of H$_2^+$) (see \refcite{Mori-Sanchez/etal:2012,Paier/etal:2010,Henderson/Scuseria:2010,Eshuis/Bates/Furche:2011} for more details).

\subsubsection{\label{sec:G2_AE}Atomization energies: the G2-I set}
One important molecular property for thermochemistry is the \textit{atomization energy}, given by
$E^\text{mol} - \sum_{i}E^\text{at}_i$ where $E^\text{mol}$ is the ground-state energy of a molecule
and $E^\text{at}_i$ that of the $i$-th isolated atom. According to this definition, the
negative of the atomization energy gives the energy cost to break the molecule into its individual atoms.
Here we examine the accuracy of RPA-based approaches for atomization energies of small molecules.
The RPA results for a set of 10 small organic molecules were reported in Furche's seminal work \cite{Furche:2001} where the underestimation of RPA for atomization energies was first observed. This benchmark set
is included in their recent review \cite{Eshuis/Bates/Furche:2011}. 
A widely accepted representative set for small organic molecules is the G2-I set \cite{Curtiss/etal:1991},
that contains 55 covalent molecules and will be used as an illustrative example here.
The RPA-type results for the G2-I set have recently been reported in the work by Paier \textit{et al.} 
\cite{Paier/etal:2010,Paier/etal:2012}.  

\begin{figure}
 \includegraphics[width=0.48\textwidth]{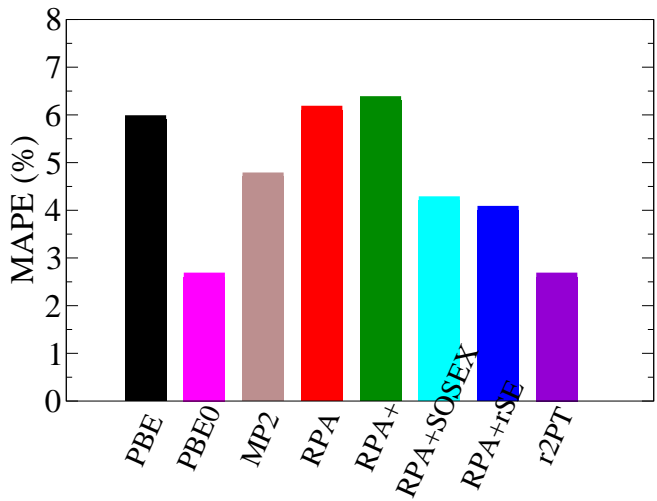}
  \caption{\label{fig:G2_MAPE}Mean absolute percentage error (MAPE) for the G2-I atomization energies
    obtained with four RPA-based approaches in addition to PBE, PBE0 and MP2. }
\end{figure}

In Fig~\ref{fig:G2_MAPE} we present in a bar graph the mean absolute percentage error (MAPE) for the 
G2-I atomization energies obtained by four RPA-based approaches in addition to GGA-PBE, the hybrid 
density-functional PBE0, and MP2. The actual values for the mean error (ME), mean absolute error (MAE),
MAPE, and maximum absolute percentage error (MaxAPE) are listed in table~\ref{Tab:G2_AE}
in Appendix~\ref{sec:appendix_tables}.  
The calculations were performed using FHI-aims \cite{Blum/etal:2009,Ren/etal:preprint} with Dunning's 
cc-pV6Z basis \cite{Dunning:1989,Wilson/vanMourik/Dunning:1996}. 
Reference data are taken from \refcite{Feller/Peterson:1999} and corrected  for zero-point energies. 
Fig~\ref{fig:G2_MAPE} and Table~\ref{Tab:G2_AE} illustrate that among the three traditional approaches, the hybrid functional PBE0 performs best, with a ME close to the ``chemical accuracy" (1 kcal/mol = 43.4 meV).  MP2 comes second, and PBE yields the largest error and shows a general trend towards overbinding. Concerning 
RPA-based approaches, standard RPA leads to ME and MAE that are even larger than the corresponding PBE 
values, with a clear trend of underbinding.  RPA+  does not improve the atomization energies 
\cite{Furche:2001}.  All this is in line with previous observations
\cite{Furche:2001,Paier/etal:2010,Paier/etal:2012,Ren/etal:preprint}. 
  As shown in the previous section, the underbinding of RPA for small molecules can be alleviated by adding the SOSEX or rSE correction. And the combination of the two (i.e. r2PT) further brings the MAE down to 3.3 kcal/mol, comparable to the corresponding PBE0 value. Whether the mechanism of this improvement can be interpreted in terms of the ``multi-center non-locality of the correlation hole" as invoked by Ruzsinszky et al. \cite{Ruzsinszky/etal:2010,Ruzsinszky/etal:2011} or not, is not yet clear at the moment.

\subsubsection{\label{sec:vdW}vdW interactions: S22 set}

As discussed above, one prominent feature of RPA is that it captures vdW interactions that are 
of paramount importance for non-covalently bonded systems. Benchmarking RPA-based methods for vdW bonded systems is an active research field \cite{Furche/Voorhis:2005,Harl/Kresse:2008,Janesko/Henderson/Scuseria:2009,Toulouse/etal:2009,Lu/Li/Rocca/Galli:2009,Li/Lu/Nguyen/Galli:2010,Zhu/etal:2010,Jansen/etal:2010,Toulouse/etal:2011,Ren/etal:2011,Eshuis/Furche:2011}. 
Here we choose the S22 test set  \cite{Jurecka/etal:2006} as the illustrating example to demonstrate the performance of RPA for non-covalent interactions. This test set contains 22 weakly bound molecular complex of different size and bonding type (7 of hydrogen bonding, 8 of dispersion bonding, and 7 of mixed nature).  Since its inception this test set has been widely adopted as the benchmark or training dataset for computational schemes that aim at dealing with non-covalent interactions \cite{Grimme/etal:2010,M06,Tkatchenko/etal:2010,vdW-DF2, Gulans/etal:2009,Sato/Nakai:2009,Klimes/etal:2010} including RPA-based approaches
\cite{Zhu/etal:2010,Ren/etal:2011,Eshuis/Furche:2011,Toulouse/etal:2011}. The consensus emerging from these studies is  as expected:  RPA improves the binding energies considerably over semilocal functionals. 

Quantitatively  the MAEs given by standard RPA  reported for S22 by different groups show an unexpected spread. Specifically Zhu \textit{et al.} \cite{Zhu/etal:2010} report an MAE of 2.79 kcal/mol, 
Ren \textit{et al.} 39 meV or 0.90 kcal/mol \cite{Ren/etal:2011}, and Eshuis and Furche 
0.41 kcal/mol \cite{Eshuis/Furche:2011}.  The latter authors investigated this issue in detail \cite{Eshuis/Furche:2012} and concluded that  the discrepancy is due to basis set incompleteness and  BSSE. Using Dunning's correlation consistent basis sets plus diffuse functions and extrapolating to the complete basis set (CBS) limit 
Eshuis and Furche obtained a MAE of 0.79 kcal/mol with 0.02 kcal/mol uncertainty \cite{Eshuis/Furche:2012}. These authors confirmed our observation that standard RPA  generally underbinds weakly bound molecules.  The basis set we have used, NAO \textit{tier} 4 plus diffuse functions  from aug-cc-pV5Z (denoted as ``\textit{tier} 4 + a5Z-d") \cite{Ren/etal:2011,Ren/etal:preprint}, yields RPA results  very close to the CBS limit. The results reported in \refcite{Ren/etal:2011} are, in our opinion, the most reliable RPA results (based on the PBE reference) for S22 so far.

\begin{figure}
\includegraphics[clip,width=0.48\textwidth]{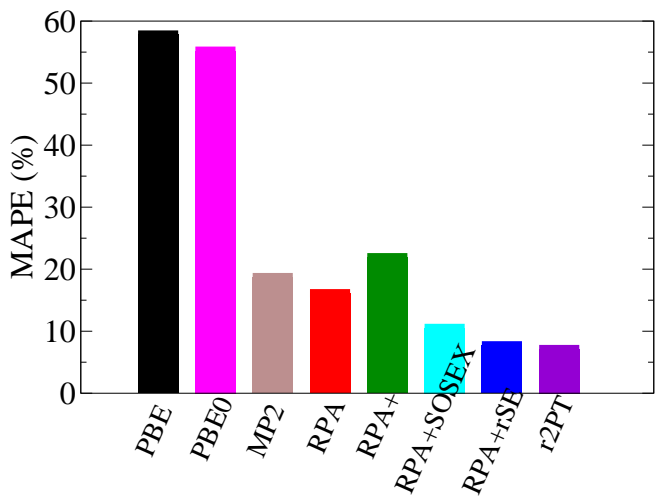}
\caption{\label{Fig:S22_MAPE}The MAPEs for the S22 test set obtained with
 RPA-based approaches in addition to PBE, PBE0, and MP2. The ``\textit{tier} 4 + a5Z-d" basis
 set was used in the calculations.}
\end{figure}

\begin{figure}
\includegraphics[clip,width=0.48\textwidth]{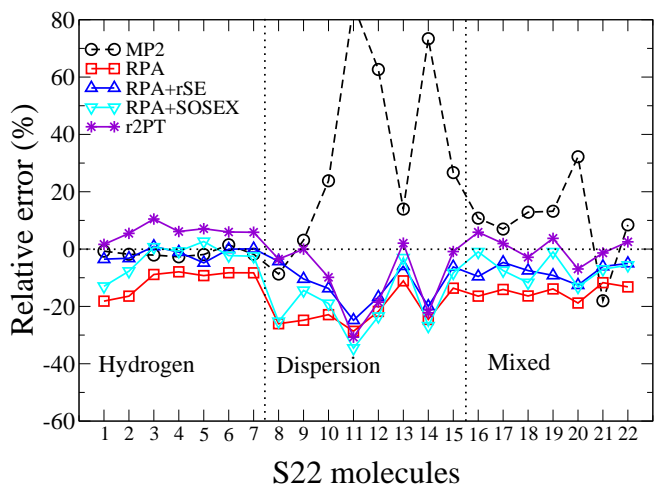}
\caption{\label{Fig:S22_relative_error}The relative errors (in $\%$) for the individual S22 molecules obtained with
 RPA-based approaches in addition to MP2. Connection lines are just guide to eyes}
\end{figure}

In Fig~\ref{Fig:S22_MAPE} the relative errors (in percentage) of five RPA-based schemes are plotted for the molecules of the S22 set. Results for PBE, PBE0, and MP2 are also presented for comparison. For MP2 and RPA-based methods, the relative errors (in percentage) for the 22 individual molecules are further demonstrated in Fig~\ref{Fig:S22_relative_error}. The reference data were obtained using CCSD(T) and properly extrapolated to the CBS limit by  Takatani \textit{et al.} \cite{Takatani/etal:2010}. MP2 and RPA results are taken from \refcite{Ren/etal:2011}. The RPA+rSE, RPA+SOSEX, and r2PT results are presented for the first time. Further details for these calculations and an in-depth discussion will be presented in a forthcoming paper \cite{Ren/etal:inpreparation}. Figure~\ref{Fig:S22_MAPE} shows that PBE and PBE0 fail drastically in this case, because these two functionals do not capture vdW interactions by construction, whereas all other methods show significant improvement. Figure~\ref{Fig:S22_relative_error} further reveals that MP2 describes the hydrogen-bonded systems very accurately, but vastly overestimates the strength of dispersion interactions, particularly for the $\pi$-$\pi$ stacking systems. Compared to MP2, RPA provides a more balanced description of all bonding types, but shows a general trend to underbind. It has been shown that this underbinding is significantly reduced by adding SE corrections \cite{Ren/etal:2011}. The renormalized SE correction presented here gives rise to a more systematic correction to RPA, as can be seen from Fig~\ref{Fig:S22_relative_error}. SOSEX shows a similar correction pattern as rSE for hydrogen bonding and mixed interactions,  but has little effect on the dispersion interaction. The r2PT scheme, that combines SOSEX and rSE corrections, overshoots for hydrogen bonding, but on average improves the description of the other two bonding types.
Figure~\ref{Fig:S22_relative_error} also reveals that $\pi$-$\pi$ stacking configurations, as exemplified by the benzene
dimer in the slip parallel geometry (\# 11), represent the most challenging case for RPA-based methods.
The relative error of RPA for this case is the largest. rSE provides little improvement, whereas SOSEX worsens 
slightly. More work is needed to understand the origin of this failure.

The detailed errors (ME, MAE, MAPE, and MaxAPE) for S22 are presented in table~\ref{Tab:S22_error} in Appendix~\ref{sec:appendix_tables}. 
Among the approaches we have investigated, RPA+rSE gives the smallest MAE for S22, while r2PT gives the smallest  MAPE. Due to space restrictions it is not possible to include the multitude of  computational schemes that have emerged in recent years for dealing with non-covalent interactions \cite{Grimme/etal:2010,M06,Tkatchenko/etal:2010,vdW-DF2,
Gulans/etal:2009,Sato/Nakai:2009,Klimes/etal:2010}. Compared to these approaches,
the RPA-based approaches presented here are completely parameter-free and systematic in the sense that they have a clear diagrammatic representation. 
Thus RPA-based approaches are expected to have a more general applicability, and may well serve as the reference for benchmarking other approaches for systems where CCSD(T) calculations are not feasible.

\subsubsection{\label{sec:bh}Reaction barrier heights}
One stringent test for an electronic structure method is its ability to predict  chemical reaction barrier heights, i.e., the energy difference between the reactants  and their transition state. This is a central quantity that dictates  chemical kinetics.  Semi-local density approximations typically underestimate barrier heights  \cite{Cohen/Mori-Sanchez/Yang:2008,Paier/etal:2010}.
RPA has already been benchmarked for barrier heights in two independent studies \cite{Paier/etal:2012,Eshuis/Bates/Furche:2011}. Both studies used the test sets of 38 hydrogen-transfer barrier heights (HTBH38) and 38 
non-hydrogen-transfer barrier heights (NHTBH38) designed by Zhao \textit{et al.} 
\cite{Zhao/Lynch/Truhlar:2004,Zhao/Nuria/Truhlar:2005} 
(together coined as BH76 in \refcite{Goerigk/Grimme:2010}). HTBH38 contains the forward and inverse
barrier heights of 19 hydrogen transfer reactions, whereas NHTBH38 contains 19 reactions
involving heavy atom transfers, nucleophilic substitutions, association, and unimolecular processes.
The reference data were obtained using the ``Weizmann-1" theory \cite{Martina/Oliveira:1999} -- 
a procedure to extrapolate the CCSD(T) results -- or by other 
``best theoretical estimates" \cite{Zhao/Nuria/Truhlar:2005}.
Paier \textit{et al.} \cite{Paier/etal:2012} presented results for standard RPA and ``beyond RPA" approaches 
based on the PBE reference, where a two-point cc-pVTZ$\rightarrow$cc-pVQZ basis-set extrapolation strategy 
is used.  In the work of Eshuis \textit{et al.} \cite{Eshuis/Bates/Furche:2011}, standard RPA results based on both PBE and TPSS \cite{Tao/etal:2003} references were presented, where the Def2-QZVP basis \cite{Weigend/Ahlrichs:2005} was used.
The RPA@PBE results for BH76 reported by both groups are very close, with an ME/MAE of -1.35/2.30 kcal/mol from the former and -1.65/3.10 kcal/mol from the latter.

\begin{figure}
\includegraphics[clip,width=0.48\textwidth]{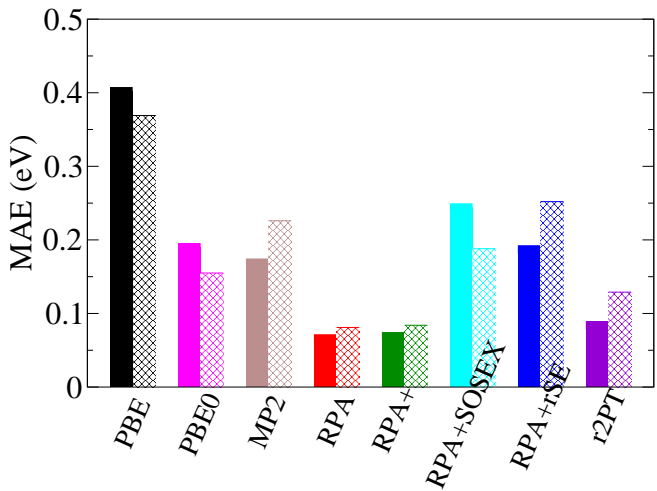}
\caption{\label{Fig:BH76_MAE}The MAEs for the HTBH38 (full bars) and NHTBH38 (hatched bars)
 test sets obtained with RPA-based approaches in addition to PBE, PBE0, and MP2.  The cc-pV6Z basis set 
 was used in the calculations.}
\end{figure}
The performance of RPA-based approaches, as well as PBE, PBE0, and MP2 for HTBH38/NHTBH38 test sets is
demonstrated by the MAE bar graph in Fig.~\ref{Fig:BH76_MAE}. The calculations were done using FHI-aims
with the cc-pV6Z basis set.  The ME, MAE, and the maximal absolute error (MaxAE) are further presented
in table~\ref{Tab:BH76_error} in Appendix~\ref{sec:appendix_tables}. In this case we do not present the
relative errors, which turn out to be very sensitive to the computational parameters due to some small 
barrier heights in the test set, and hence cannot be used as a reliable measure of
the performance of the approaches examined here.
 Compared to the results reported in \refcite{Paier/etal:2012}, besides a different basis set (cc-pV6Z instead of 
cc-pVTZ$\rightarrow$cc-pVQZ extrapolation), we also used the refined rSE correction  (see discussion in Section \ref{sec:rSE_SOSEX}) for the  RPA+rSE and r2PT results in Tab.~\ref{Tab:BH76_error}, which gives slightly better results in this case.

On average, PBE  underestimates the reaction barrier heights substantially,  a feature that is well-known for GGA functionals. The hybrid PBE0 functional reduces both the ME and MAE  by more than a factor of two. However, the remaining error is still sizable. Standard RPA performs magnificently and shows a significant improvement over PBE0. The performance of RPA+ is again very similar to standard RPA. As already noted in \refcite{Paier/etal:2012}, both the  rSE and the SOSEX corrections deteriorate the performance of RPA. This is somewhat disappointing, and highlights the challenge for designing simple, generally more accurate corrections to RPA.  Fortunately, the errors 
of rSE and SOSEX are now in the opposite direction, and largely canceled out when combining the two schemes.
Indeed, the AEs and MAEs of r2PT are not far from their RPA counterparts, although the individual errors
are more scattered in r2PT as manifested by the larger MaxAE.

	\subsection{\label{sec:solid}Crystalline solids}
		Crystalline solids are an important domain for RPA based approaches, in particular because the quantum chemical hierarchy of benchmark approaches cannot easily be transfered to periodic systems.  Over the years RPA calculations have been performed for a variety of systems such as Si  \cite{Miyake/etal:2002,Gonzalez/Fernandez/Rubio:2007,Nguyen/deGironcoli:2009}, Na \cite{Miyake/etal:2002}, h-BN \cite{Marini/Gonzalez/Rubio:2006},  NaCl \cite{Gonzalez/Fernandez/Rubio:2007}, rare-gas solids \cite{Harl/Kresse:2008}, graphite \cite{Harl/Kresse:2009,Lebegue/etal:2010}, and benzene crystals \cite{Lu/Li/Rocca/Galli:2009}.  The most systematic benchmark study of RPA for crystalline solids was conducted by Harl, Schmika, and Kresse \cite{Harl/Kresse:2009,Harl/Schimka/Kresse:2010}. These authors reported ``technically converged" calculations using their VASP code and the projector augmented plane wave method  for atomization energies, lattice constants, and bulk moduli of 24 representative crystals, including ionic compounds (MgO, LiF, NaF, LiCl, NaCl),
semiconductors (C, Si, Ge, SiC, AlN, AlP, AlAs, GaN, GaP, GaAs, InP, InAs, InSb), and 
metals (Na, Al, Cu, Rh, Pd, Ag). The error analysis of their RPA and RPA+ results, based
on a PBE reference, as well as the LDA and PBE results are presented in Tab.~\ref{Tab:solid}.
\begin{table}
  \caption{\label{Tab:solid}ME, MAE, MAPE (\%), and MaxAPE (\%) 
    for the atomization energies (in eV/atom), lattice constants (in \AA),
    and bulk moduli (in Gpa)  of 24 crystalline solids.  Results are taken from \refcite{Harl/Schimka/Kresse:2010}.  
    The experimental atomization energies \refcite{NIST_solids} are corrected for temperature effect (based
    on thermochemical correction data) \cite{Paier/Marsman/Kresse:2007} and zero-point vibrational energy. 
    The experimental lattice constants have been corrected for anharmonic expansion effects.}
  \begin{tabular}{lcccc}
    \hline\hline \\[-2ex]
    & \multicolumn{4}{c}{Atomization energies} \\[0.2ex]
    \cline{2-5} \\[-1.5ex]
    & ME (eV) & MAE (eV) & MAPE (\%) & MaxAPE (\%) \\[0.2ex]
    \cline{2-5} \\[-1.5ex]
   LDA & $-0.74$  & 0.74 & 18.0 &  32.7 \\
   PBE & ~ 0.15 & 0.17 & ~4.5 &   15.4 \\
   RPA & ~ 0.30  & 0.30 & ~7.3  &  13.5 \\ 
   RPA+ & ~ 0.35 & 0.35 & ~8.7 &  15.0 \\
    \hline \\[-1.5ex]
    & \multicolumn{4}{c}{Lattice constants} \\[0.2ex]
    \cline{2-5} \\[-1.5ex]
    & ~ME (\AA)~ & ~MAE (\AA)~ & MAPE (\%) & MaxAPE (\%) \\[0.2ex]
    \cline{2-5} \\[-1.5ex]
   LDA & $-0.045$ & 0.045 & 1.0 & 3.7 \\
   PBE & ~ 0.070 &   0.072 & 1.4 & 2.7 \\
   RPA & ~ 0.016 &  0.019 &  0.4 & 0.9 \\
   RPA+ & ~ 0.029 & 0.030 &  0.6 & 1.1  \\
    \hline \\[-1.5ex]
    & \multicolumn{4}{c}{Bulk Moduli} \\[0.2ex]
    \cline{2-5} \\[-1.5ex]
    & ~ME (GPa)~ & ~MAE (GPa)~ & MAPE (\%) & MaxAPE (\%) \\[0.2ex]
    \cline{2-5} \\[-1.5ex]
   LDA & ~ 9 & 11 & ~9.6 & 31.0 \\
   PBE & $-11$ &  11 & 10.7 & 23.7 \\
   RPA & $-1$~ &  ~4 & ~3.5 & 10.0 \\
   RPA+ & $-3$~ & ~5 & ~3.8 & 11.4  \\
   \hline\hline
  \end{tabular}
\end{table}
As is clear from Tab.~\ref{Tab:solid}, the  RPA lattice constants and bulk moduli
 are better than in LDA and PBE. The atomization
energies, however, are systematically underestimated in RPA, and the MAE
in this case is even larger than that of PBE. This behavior is very similar to that for the atomization energies in the G2 set discussed above.  Harl \textit{et al.} also observed that the error of RPA does not grow  when going to heavier atoms, or open-shell systems in contrast to LDA or PBE \cite{Harl/Schimka/Kresse:2010}. The RPA+ results are in general slightly worse than those of standard RPA.

The performance of RPA has not been extensively benchmarked for vdW-bound solids. However, judging
from the studies on h-BN \cite{Marini/Gonzalez/Rubio:2006}, rare-gas 
crystals \cite{Harl/Kresse:2008}, benzene crystal \cite{Lu/Li/Rocca/Galli:2009}, and
graphite \cite{Harl/Kresse:2009}, standard RPA does an excellent job regarding the equilibrium lattice
constants and cohesive energies, whereas semi-local DFT fails miserably, yielding typically 
too weak binding and too large lattice constants. LDA typically gives a finite binding 
(often overbinding) for weakly bonded solids, but its performance varies significantly from system
to system, and cannot be trusted in general, as the functional by construction does not contain
the necessary physics to reliably describe this phenomenon.  In a recent work \cite{Lebegue/etal:2010}, 
it was shown that RPA also reproduces the correct $1/d^3$ asymptotics between graphite
layers as analytically predicted by Dobson \textit{et al.} \cite{Dobson/White/Rubio:2006}.
This type of behavior can neither be described by LDA, GGAs, nor by hybrid functionals.

For solids attempts have also been made to go beyond the standard RPA. For a smaller
test set of 11 insulators, Paier \textit{et al.} \cite{Paier/etal:2012} showed 
that adding SOSEX corrections to RPA  the MAE of atomization energies  
is reduced from 0.35 eV/atom to 0.14 eV/atom. By replacing the non-self-consistent
HF energy by its self-consistent counterpart, which mimicks the effect of adding single 
excitation corrections \cite{Ren/etal:2011}, reduces the MAE further to 0.09 
eV/atom \cite{Paier/etal:2012}. Thus the trend in periodic insulators is again 
in line with what has been observed for molecular atomization energies. The effects of SOSEX
and rSE corrections for metals, and for other properties such as the lattice constants 
and bulk moduli have not been reported yet.


	\subsection{\label{sec:surf}Adsorption at surfaces}
The interaction of atoms and molecules with surfaces plays a significant role in many phenomena in surface science and 
for industrial applications. In practical calculations,  the super cells needed to model the surfaces are large and a good electronic structure approach has to give a balanced description for both the solid and the adsorbate, as well as the interface between the two. Most approaches today perform well for either the solid or the isolated adsorbate (e.g. atoms, molecules, or clusters), but not for the combined system, or are computationally too expensive to be applied to large super cells. This is an area where we believe RPA will prove to be advantageous.

The systems to which RPA has been applied  include Xe and  3,4,9,10-perylene-tetracarboxylic acid dianhydride (PTCDA) adsorbed on Ag(111)  \cite{Rohlfing/Bredow:2008}; CO on Cu(111) \cite{Ren/Rinke/Scheffler:2009,Harl/Kresse:2009}
and other noble/transition metal surfaces \cite{Schimka/etal:2010}; benzene on Ni(111) \cite{Schimka/etal:2010}, Si(001) \cite{Kim/etal:2012}, and the graphite surface\cite{Ma/etal:2011}; and graphene on Ni(111)  \cite{Mittendorfer/etal:2011,Olsen/etal:2011}, and Cu(111), Co(0001) surfaces \cite{Olsen/etal:2011}. 
In all these applications,  RPA has been very  successful.

\begin{figure}
\includegraphics[clip,width=0.45\textwidth]{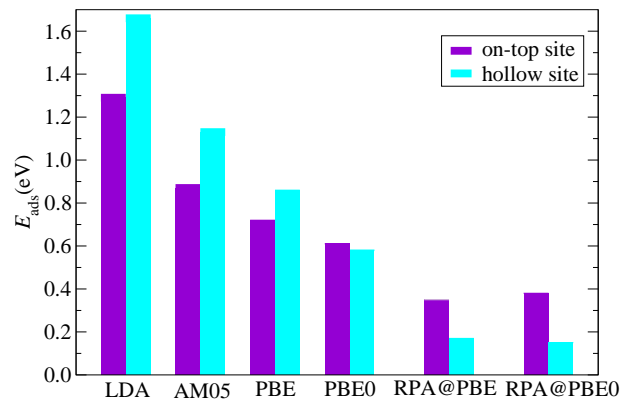}
\caption{\label{Fig:CO_ads}Adsorption energies for CO adsorbed at the on-top and fcc hollow sites of 
  the Cu(111) surface as obtained using LDA, AM05, PBE, PBE0, and RPA. RPA results are presented
  for both PBE and PBE0 references, and  they differ very little.}
\end{figure}

To illustrate how RPA works for an adsorbate system, here we briefly describe the RPA study of CO@Cu(111) following 
\refcite{Ren/Rinke/Scheffler:2009}. The work was motivated by the so-called ``CO adsorption puzzle" -- LDA and 
several GGAs predict the wrong adsorption site for CO adsorbed on several noble/transition metal surfaces at low
coverage \cite{Feibelman:2001}.  For instance, for the (111) surface of Cu and Pt DFT within local/semi-local approximations erroneously favor the threefold-coordinated hollow site, whereas experiments clearly show that the singly-coordinated on-top site is the energetically most stable site \cite{Steininger/Lehwald/Ibach:1982,Blackman/etal:1988}. This posed a severe challenge to the first-principles modeling of molecular adsorption problems and the question arose, at what level of approximation can the correct physics be recovered. 
In our study, the Cu surface was modeled using systematically increasing Cu clusters cut out of the Cu(111) surface. 
Following a procedure proposed by
Hu \textit{et al.} \cite{Hu/Reuter/Scheffler:2007}, the RPA adsorption energy was obtained 
by first converging its difference to the PBE values with respect to cluster size, and then adding the converged difference to the periodic  PBE results. The RPA adsorption energies for both the on-top and fcc (face centered cubic) hollow sites are presented in \Fref{Fig:CO_ads}, together with the results from LDA, AM05 \cite{Armiento/Mattsson:2005}, PBE, and the hybrid PBE0 functional.
\Fref{Fig:CO_ads} reveals what happens in the CO adsorption puzzle when climbing the so-called Jacob's ladder in DFT \cite{Perdew/Schmidt:2001} 
--- going from the first two rungs (LDA and GGAs) to the fourth (hybrid functionals), and finally to the fifth rung (RPA and other functionals that explicitly depend on unoccupied KS states). Along the way the magnitude of the adsorption energies
on both sites are reduced, but the effect is more pronounced for the fcc hollow site. The correct energy ordering is already restored at the PBE0 level, but the energy between the two sites is too small. RPA not only gives the correct adsorption site, but also produces a reasonable adsorption energy  difference of 0.22 eV, consistent with experiments.
This result was later confirmed by the periodic RPA calculations of Harl and Kresse in \refcite{Harl/Kresse:2009}, with only small numerical differences arising from the different implementations and different convergence strategy.

The work of Rohlfing and Bredow on Xe and PTCDA adsorbed at Ag(111) surface represents the first
RPA study regarding surface adsorption problems, where the authors explicitly demonstrated that RPA 
yields the expected $-C_3/(d-d_0)^3$ behavior for large molecule-surface separations $d$. 
Schimka \textit{et al.} extended the RPA benchmark studies of the CO adsorption problem to more surfaces 
\cite{Schimka/etal:2010}. They found that RPA is the only approach so far that gives both good adsorption energies as well as surface energies. GGAs and hybrid functionals at most yield either good surface energies, or adsorption energies, but never both. 
G\"{o}ltl and Hafner investigated the adsorption of small alkanes in Na-exchanged chabazite using RPA and several other approaches. They found that the ``hybrid RPA" scheme, as proposed in \refcite{Ren/etal:2011} and further examined in \refcite{Paier/etal:2012},
provides the most accurate description of the system compared to the alternatives e.g. DFT-D \cite{Grimme:2004} and vdW-DF \cite{Dion/etal:2004}. More recently RPA was applied to the adsorption of benzene on the 
Si(001) surface by Kim \textit{et al} \cite{Kim/etal:2012}, graphene on the Ni(111) surface by
Mittendorfer \textit{et al} \cite{Mittendorfer/etal:2011} and by Olsen \textit{et al.} \cite{Olsen/etal:2011},
and additionally graphene on Cu(111) and Co(0001) surfaces by the latter authors. In all these studies, RPA 
is able to capture the delicate balance between covalent and dispersive interactions, and yields quantitatively
reliable results.  We expect RPA to become increasingly more important in surface science with increasing computer power and more efficient implementations.

\section{\label{sec:summary}Discussion and Outlook}
        RPA is an important concept in physics and has a more than 50 year old history. Owing to its rapid development  in recent years, RPA has shown great promise as a powerful first-principles electronic-structure method with significant implications for quantum chemistry, computational physics, and materials science in the  foreseeable future. 
The rise of the RPA method in electronic structure theory, and its recent generalization to r2PT, were borne out by realizing that traditional DFT functionals (local and semi-local approximations) are encountering noticeable accuracy and reliability limits and that hybrid  density functionals are not sufficient to overcome them. 
With the rapid development of computer hardwares and algorithms,  it is not too ambitious to expect RPA-based approaches to become (or at least to inspire) main-stream electronic-structure methods in computational materials science and engineering in the coming decades.  At this point it would be highly desirable if the community would start to build up benchmark sets for materials
science akin to the ones in quantum chemistry (e.g. G2 \cite{Curtiss/etal:1991} or S22 \cite{Jurecka/etal:2006}). These should include prototypical bulk crystals, surfaces, and surface adsorbates and would aid the development of RPA-based approaches.

As an outlook, we would like to indicate several directions for future developments of RPA-based methods.
\begin{itemize}
\item[i.] \emph{Improved accuracy}:  Although RPA does not suffer from the well-documented pathologies of LDA and GGAs, its  quantitative accuracy is not always what is desired, in particular for atomization energies. To improve on this and to make RPA worth its computational effort, further corrections to RPA are necessary. To be useful in practice, these should not increase the computational cost significantly.  The r2PT approach as presented in \refsec{sec:beyond} and benchmarked in \refsec{sec:appl} is one example of this kind. More generally the aim is to develop RPA-based computational schemes that are close in accuracy to CCSD(T), but come at a significantly reduced numerical cost. More work can and should be done  along this direction.

\item[ii.] \emph{Reduction of the computational cost}: The major factor that currently prevents the widespread
use of RPA in materials science is its high numerical cost compared to traditional DFT methods. The state-of-the-art implementations still have an $O(N^4)$ scaling, as discussed in \refsec{sec:comp}. To enlarge the domain of RPA applications, a reduction of this scaling behavior will be highly desirable.  Ideas can be borrowed from $O(N)$ methods \cite{Ochsenfeld/etal:2007} developed in quantum-chemistry (in particular in the context of MP2) or compression techniques applied in the $GW$ context \cite{Foerster/etal:2011}.

\item[iii.] \emph{RPA forces}: For a ground-state method, one crucial component that is still missing in RPA are atomic forces.  Relaxations of atomic geometries that are common place in DFT and that make DFT such a powerful method are currently not possible with RPA or at least have not been demonstrated yet. An efficient realization of RPA forces would therefore extend its field of application to  many more interesting and important materials science problems.

\item[iv.] \emph{Self-consistency}: Practical RPA calculations are predominantly done in a post-processing manner, in which single-particle orbitals from KS or generalized KS calculations are taken as input for a one-shot RPA calculation. This introduces undesired uncertainties, although the starting-point dependence  is often not very pronounced, if one restricts the input to KS orbitals. A self-consistent RPA approach can be defined within KS-DFT via the optimized effective potential method \cite{Kuemmel/Kronik:2008}, and has already been applied in a few instances \cite{Kotani:1998,Gruening/Marini/Rubio:2006,Hellgren/Barth:2007,Hellgren/etal:2012,Verma/Bartlett:2012}.  However, in its current realizations self-consistent RPA is numerically very challenging, and a more practical, robust and numerically more efficient procedure will  be of great interest.

\end{itemize}

With all these developments, we expect RPA and its generalizations will play an increasingly important role in computational materials science in the near future.

\begin{appendix}
\section{RI-RPA implementation in FHI-aims}
\label{sec:appendix_RI-FHI-aims}
In this section we will briefly describe how RPA is implemented in the FHI-aims code \cite{Blum/etal:2009} using the resolution-of-identity (RI) technique. More details can be found in \refcite{Ren/etal:preprint}. For a different formulation of RI-RPA see 
\refcite{Eshuis/Yarkony/Furche:2010,Eshuis/Bates/Furche:2011}. We start with the expression for the RPA correlation energy in Eq.~\refeq{Eq:Ec_RPA}, which can be formally expanded in a Taylor series,
 \begin{equation}
   E^\text{RPA}_\text{c} = -\frac{1}{\pi} \int_0^\infty d\omega \sum_{n=2}^\infty \frac{1}{2n}
              \text{Tr}\left[\left( \chi^0(i\omega)v \right)^n \right].
   \label{Eq:Ec_rpa_Taylor}
 \end{equation}
Applying RI to RPA in this context means to represent both $\chi^0(i\omega)$ and $v$ in an appropriate
auxiliary basis set. Eq.~\refeq{Eq:Ec_rpa_Taylor} can then be cast into a series of matrix operations. To achieve this we perform the following RI expansion
  \begin{equation}
   \psi_i^\ast(\bfr)\psi_j(\bfr) \approx \sum_{\mu=1}^{N_\text{aux}} C_{ij}^\mu P_\mu(\bfr)\, ,
   \label{Eq:RI_expansion}
  \end{equation}
where $P_\mu(\bfr)$ are auxiliary basis functions, $C_{ij}^{\mu}$ are the expansion coefficients, and
$N_\text{aux}$ is the size of the auxiliary basis set. Here $C$
serves as the transformation matrix that reduces the rank of all matrices  from $N_\text{occ}*N_\text{vir}$
to $N_\text{aux}$, with $N_\text{occ}$, $N_\text{vir}$ and $N_\text{aux}$ being the number of 
occupied single-particle orbitals, unoccupied (virtual) single-particle orbitals, and auxiliary basis functions, 
respectively. The determination of the $C$ coefficients is not unique, but depends on the underlying metric. 
In quantum chemistry the ``Coulomb metric" is the standard choice where the $C$ coefficients are determined by minimizing
the Coulomb repulsion between the residuals of the expansion in Eq.~\refeq{Eq:RI_expansion} (for details see 
\refcite{Ren/etal:preprint} and references therein). In this so-called ``RI-V" approximation, the $C$ coefficients are
given by
  \begin{equation}
      C_{ij}^{\mu} = \sum_{\nu}(ij|\nu)V_{\nu\mu}^{-1} \, ,
      \label{Eq:V_coefficient}
  \end{equation}
where
  \begin{equation}
     (ij|\nu)=\iint \frac{\phi_i(\bfr)\phi_j(\bfr)P_\nu(\bfrp)}{|\bfr-\bfrp|} d\bfr d\bfrp \, ,
  \end{equation}
and 
\begin{equation}
  V_{\mu\nu}=\int \frac{P_\mu(\bfr) P_\nu(\bfrp)}{|\bfr-\bfrp|}
     d\bfr d\bfrp \, .
   \label{Eq:coulomb_matrix}
\end{equation}

In practice, sufficiently 
accurate auxiliary basis set can be constructed  such that $N_\text{aux} \ll N_\text{occ}*N_\text{vir}$, thus reducing the
computational effort considerably. A practically accurate and efficient way of constructing auxiliary basis set
$\{P_\mu(\bfr)\}$ and their associated $\{C_{ij}^\mu\}$ for atom-centered basis functions of general shape
has been presented in \refcite{Ren/etal:preprint}.

Combining Eq.~\refeq{Eq:indep_response} with \refeq{Eq:RI_expansion} yields
 \begin{align}
   \chi^0(\bfr,\bfrp, i\omega)& = \sum_{\mu\nu} \sum_{ij} \frac{(f_i-f_j)C_{ij}^\mu C_{ji}^\nu}
    {\epsilon_i - \epsilon_j - i\omega}  P_\mu(\bfr)P_\nu(\bfrp) \nonumber \\
     & = \sum_{\mu\nu} \chi^0_{\mu\nu}(i\omega) P_\mu(\bfr)P_{\nu}(\bfrp),
 \end{align}
 where 
 \begin{equation}
   \chi^0_{\mu\nu}(i\omega) = \sum_{ij} \frac{(f_i-f_j)C_{ij}^\mu C_{ji}^\nu}{\epsilon_i - \epsilon_j - i\omega}.
  \label{Eq:chi_0_matr}
 \end{equation}
Introducing the Coulomb matrix
 \begin{equation}
   V_{\mu\nu} = \iint d\bfr\bfrp P_\mu(\bfr) v(\bfr,\bfrp)P_{\nu}(\bfrp)\, ,
  \label{Eq:V_matr}
 \end{equation}
we obtain the first term in \refeq{Eq:Ec_rpa_Taylor}
  \begin{align}
   E_\text{c}^{(2)}  = & -\frac{1}{4\pi} \int_0^\infty d\omega \int\cdots\int d\bfr d\bfr_1 d\bfr_2 d\bfrp \times 
              \nonumber  \\
            &  \chi^0(\bfr,\bfr_1)
              v(\bfr_1,\bfr_2) \chi^0(\bfr_2,\bfrp) v(\bfrp,\bfr) \nonumber \\
        = & -\frac{1}{4\pi} \int_0^\infty d\omega  \sum_{\mu\nu,\alpha\beta} \chi^0_{\mu\nu}(i\omega)
           V_{\nu\alpha} \chi^0_{\alpha\beta}(i\omega) V_{\beta,\mu} \nonumber \\
        = & -\frac{1}{4\pi} \int_0^\infty d\omega \text{Tr}\left[\left(\chi^0(i\omega)V\right)^2\right]. 
  \end{align}
This term corresponds to the 2nd-order direct correlation energy also found in MP2. Similar equations hold for the higher order terms in \refeq{Eq:Ec_rpa_Taylor}. This suggests that the trace operation
in Eq.~\refeq{Eq:Ec_RPA} can be re-interpreted as a summation over auxiliary basis function indices, namely, 
$\text{Tr}\left[AB\right] = \sum_{\mu\nu}A_{\mu\nu}B_{\nu\mu}$, provided that $\chi^0(\bfr,\bfrp,i\omega)$ 
and $v(\bfr,\bfrp)$
 are represented in terms of a suitable set of auxiliary basis functions. Equations (\ref{Eq:Ec_RPA}), 
(\ref{Eq:chi_0_matr}), and (\ref{Eq:V_matr}) constitute a practical scheme for RI-RPA.
In this implementation, the most expensive step is Eq.~\refeq{Eq:chi_0_matr} for the construction of the independent
response function, scaling as $N_\text{aux}^2N_\text{occ}N_\text{vir}$. We note that the same is true for 
standard plane-wave implementations \cite{Harl/Schimka/Kresse:2010} where $N_\text{aux}$ corresponds to the number 
of plane waves used to expand the response function. In that sense RI-based local-basis function implementations are very similar to plane-wave-based or LAPW-based implementations,  where the plane waves themselves or
the mixed product basis serve as the auxiliary basis set.

\section{Error statistics for G2-I, S22, and NHBH38/NHTBH38 test sets}
\label{sec:appendix_tables}
Tables~\ref{Tab:G2_AE}, \ref{Tab:S22_error}, and \ref{Tab:BH76_error} present a more detailed error analysis for the
G2-I, S22, and NHBH38/NHTBH38 test sets. Given are the mean error (ME), the mean absolute error (MAE), the mean absolute percentage error (MAPE), the maximum absolute percentage error (MaxAPE), and the maximum absolute error (MaxAE).
\begin{table}[h!]
\caption{\label{Tab:G2_AE} ME (in eV), MAE (in eV), MAPE (\%), MaxAPE(\%) for 
 atomization energies of the G2-I set obtained  with four RPA-based approaches in addition to PBE, PBE0 and MP2. A negative ME indicates overbinding (on average) and a positive ME underbinding. The cc-pV6Z basis set was used.}
\begin{tabular*}{0.45\textwidth}{lr@{.}lr@{.}lr@{.}lr@{.}lr@{.}l}
\hline\hline  \\[-2ex]
 \multicolumn{1}{c}{}& \multicolumn{2}{c}{~~ME~~} & \multicolumn{2}{c}{~~MAE~~}& \multicolumn{2}{c}{MAPE~~} & 
     \multicolumn{2}{c}{MaxAPE~~} \\[0.5ex]
\hline \\[-2ex]
     PBE    &    -0&28     &     0&36     &      5&9     &     38&5  \\
    PBE0    &     0&07     &     0&13     &      2&6     &     20&9  \\
     MP2    &    -0&08     &   ~~~0&28    &   ~~~4&7     &  ~~~24&6  \\
     RPA    &     0&46     &     0&46     &      6&1     &     24&2   \\
    RPA+    &     0&48     &     0&48     &      6&3     &     24&9   \\
RPA+SOSEX    &     0&22     &     0&25     &      4&2     &     36&7  \\
 RPA+rSE    &     0&30     &     0&31     &      4&0     &     22&5   \\
    r2PT    &     0&07     &     0&14     &      2&6     &     24&9   \\[0.5ex] %
\hline\hline
\end{tabular*}
\end{table}
\begin{table}[h!]
\caption{\label{Tab:S22_error}ME (in meV), MAE (in meV), MAPE (\%), and MaxAPE(\%)
for the S22 test set \cite{Jurecka/etal:2006} obtained with five RPA-based approaches in addition to PBE, PBE0, and MP2  obtained with FHI-aims.  The basis set ``\textit{tier} 4 + a5Z-d" \cite{Ren/etal:preprint} was used in all calculations.} 
\begin{tabular*}{0.45\textwidth}{lp{1mm}cp{2mm}cp{1mm}cc}
\hline\hline
 & &  ME  & & MAE  &  & MAPE & MaxAPE \\
\hline
PBE & & 116.2 & & 116.2 & & 57.8 & 170.3 \\  
PBE0 & & 105.7 & & 106.5 & & 55.2 & 169.1 \\
MP2 & & -26.5 & & 37.1 & & 18.7 & ~85.1 \\
RPA & & ~37.8 & & 37.8 & & 16.1 & ~28.7 \\
RPA+ & & ~51.2 & & 51.2 & & 21.9 & ~39.4 \\
RPA+rSE & & ~14.1 & & 14.8 & & ~7.7 & ~24.8 \\
RPA+SOSEX & & ~15.4 & & 18.0 & & 10.5 & ~34.6 \\
r2PT & & ~-8.4 & & 21.0 & & ~7.1 & ~30.7 \\
\hline\hline
\end{tabular*}
\end{table}
\begin{table}[h!]
\caption{\label{Tab:BH76_error}ME, MAE, and MaxAE (in eV)
for the HTBH38 \cite{Zhao/Lynch/Truhlar:2004} and NHTBH38 \cite{Zhao/Nuria/Truhlar:2005} test sets
obtained with four RPA-based approaches in addition to PBE, PBE0, and MP2, as obtained using FHI-aims.
The cc-pV6Z basis set was used in all calculations. Negative ME indicates an underestimation of
the barrier height on average.}
\begin{tabular*}{0.48\textwidth}{lcccp{0.2cm}ccc}
\hline\hline \\[-2ex]
 & \multicolumn{3}{c}{HTBH38} & & \multicolumn{3}{c}{NHTBH38} \\[0.2ex]
     \cline{2-8}   \\[-2ex]
 &   ME  &  MAE  &  MaxAE & & ME  &  MAE  &  MaxAE  \\[0.2ex]
\hline
    PBE    &    -0.399    &     0.402    &     0.863    &    &    -0.365    &     0.369    &     1.320   \\
    PBE0    &    -0.178    &     0.190    &     0.314    &    &    -0.134    &     0.155    &     0.609   \\
     MP2    &    ~0.131    &     0.169    &     0.860    &    &    ~0.215    &     0.226    &     1.182   \\
     RPA    &    ~0.000    &     0.066    &     0.267    &    &    -0.065    &     0.081    &     0.170   \\
    RPA+    &    ~0.005    &     0.069    &     0.294    &    &    -0.068    &     0.084    &     0.168   \\
 RPA+rSE    &    -0.170    &     0.187    &     0.809    &    &    -0.251    &     0.252    &     0.552   \\
RPA+SOSEX   &    ~0.243    &     0.244    &     0.885    &    &    ~0.185    &     0.188    &     0.781   \\
    r2PT    &    ~0.072    &     0.084    &     0.453    &    &    -0.001    &     0.129    &     0.432   \\[0.2ex]
\hline\hline
\end{tabular*}
\end{table}

\end{appendix}
\bibliography{./CommonBib,./JoasBib}
\end{document}